\pdfoutput=1
\documentclass[
 reprint,
 amsmath,amssymb,
 aps,
 pre,
 floatfix,
 pgfplots,
 lengthcheck,
]{revtex4}
\usepackage{graphicx}
\usepackage{epstopdf}
\usepackage{bm}
\let\oldhat\hat
\renewcommand{\vec}[1]{\mathbf{#1}}
\renewcommand{\hat}[1]{\oldhat{\mathbf{#1}}}

\begin{document}
\title{Fusion in a magnetically-shielded-grid inertial electrostatic confinement device}
\author{John Hedditch}
\email{j.hedditch@gmail.com}
\author{Richard Bowden-Reid}
\email{rbow3948@physics.usyd.edu.au}
\author{Joe Khachan}
\email{joe.khachan@sydney.edu.au}

\affiliation{School of Physics, The University of Sydney, NSW
2006, Australia}

\date{\today}
\begin{abstract}
Theory for a gridded inertial electrostatic confinement (IEC) fusion
system is presented that shows a net energy gain is possible if
the grid is magnetically shielded from ion impact. A simplified
grid geometry is studied, consisting of two negatively-biased coaxial current-carrying rings,
oriented such that their opposing magnetic fields produce a spindle cusp.
Our analysis indicates that better than break-even performance is possible
even in a deuterium-deuterium system at bench-top scales.
The proposed device has the unusual property that it
can avoid both the cusp losses of traditional magnetic fusion
systems and the grid losses of traditional IEC configurations.
\end{abstract}


\maketitle

\section{Introduction}

\begin{figure}[h]
\vspace{-5mm}
\includegraphics[width=7.5cm, height=7.4cm]{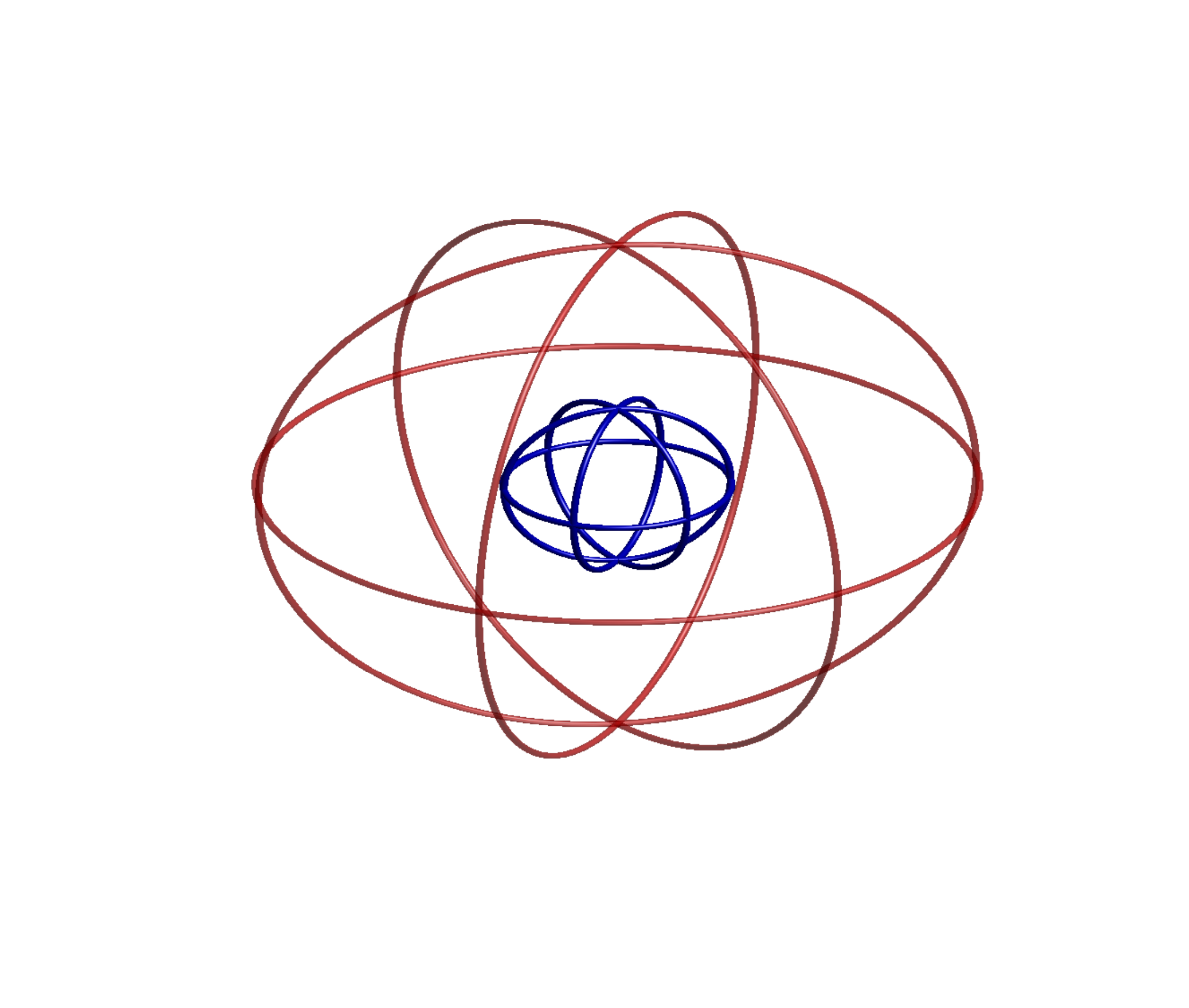}
\vspace{-10mm}
\caption{\label{fig1}A schematic diagram of a spherical gridded
IEC device, where the inner electrode is the cathode.}
\end{figure}

Inertial Electrostatic Confinement (IEC) is a method of producing
nuclear fusion by trapping and heating ions in a spherical or
quasi-spherical electrostatic potential well~\cite{farnsworth1, farnsworth2, hirsch1, hirsch2}. Various schemes
have been implemented~\cite{elmore1,
lavrentev1, miley1, barnes1, park1}, the most common of which
- known as gridded IEC - employs a concentric spherical cathode
and anode arrangement as shown in Fig.~\ref{fig1}. The
inner electrode, shown in blue, is the cathode and must be
highly transparent to converging ions.

Although one can apply a sufficiently large
electric field to accelerate ions to fusion-relevant energies,
many recirculating ions are intercepted by the
cathode grid, resulting in significant heating loss.
At bench-top scale, the
losses are at least five orders of magnitude larger than
the fusion power produced, even when the devices are
operated in the so-called ``star-mode''.
Nevertheless, these devices have found application as
neutron sources~\cite{masuda1, miley2}.

Other schemes have been proposed~\cite{
tuft1,  bandara1} in attempts to circumvent grid losses.
Such methods aim to replace the grid with a
{\it virtual cathode} formed through the confinement of electrons.
Specifically, applying a positive bias to the inner grid results
in a central electrons focus and consequent virtual cathode. A
potential well is subsequently formed between the anode grid and
the electron-rich core. Although relatively
deep potential wells have been formed and significant ion heating
obtained in this way, electron loss to the grid limits the
attainable efficiency.

Alternatively, a virtual cathode can be formed through the trapping of
electrons in a three dimensional magnetic cusp system, known as a
Polywell~\cite{krall1, carr1, carr2, carr3, cornish1}. This
consists of an arrangement of current loops that are positioned on
the faces of a regular polyhedron, most commonly a cube. The currents are applied so that a
magnetic null is obtained at the center of the device. Electrons
introduced into this device are trapped by the magnetic cusp
configuration for sufficient time as to form a
virtual cathode. Inefficiencies arise when electrons are scattered
into the loss cones of the face, edge, and corner cusps, allowing
them to escape the trap.

In this paper, we present a new approach to the realisation
of nuclear fusion in IEC devices. We combine IEC and magnetic
cusp confinement to yield a hybrid electromagnetic reactor. Ion
confinement and heating is provided through the electric field,
whilst the magnetic field serves both to enhance electron confinement
and to shield the (cathode) grid from ion impact. Earlier attempts
to combine magnetic and electrostatic confinement~\cite{dolan1} did
not operate in this fashion.

In the following sections we derive expressions for energy gain
and loss mechanisms for such a magnetically-shielded-grid (MSG)
fusion system. Energy gain comes from fusion, whilst
particle losses through
magnetic cusps, interspecies energy transfer, ion and electron
losses, and bremsstrahlung contribute to energy loss.

Notwithstanding a discouraging prediction of extremely poor
performance from an early exploratory paper by Rider~\cite{rider1},
significant net energy gain is predicted for a judicious choice of the
potential applied to the grid, the current in the field coils and
the energy of the interacting species. This is true even
for small-scale devices and there is little reason to presume
it would not also hold true for exotic fuels such as proton-Boron
or Helium-3, as the $Z^2$  increase in Bremsstrahlung for these
reactions would be compensated to some degree by the increased energy
per fusion event.

\section{Derivation of energy gain and loss expressions}
\subsection{Device geometry and operation}
\begin{figure}[h]
\vspace{-5mm}
\includegraphics[width=\columnwidth]{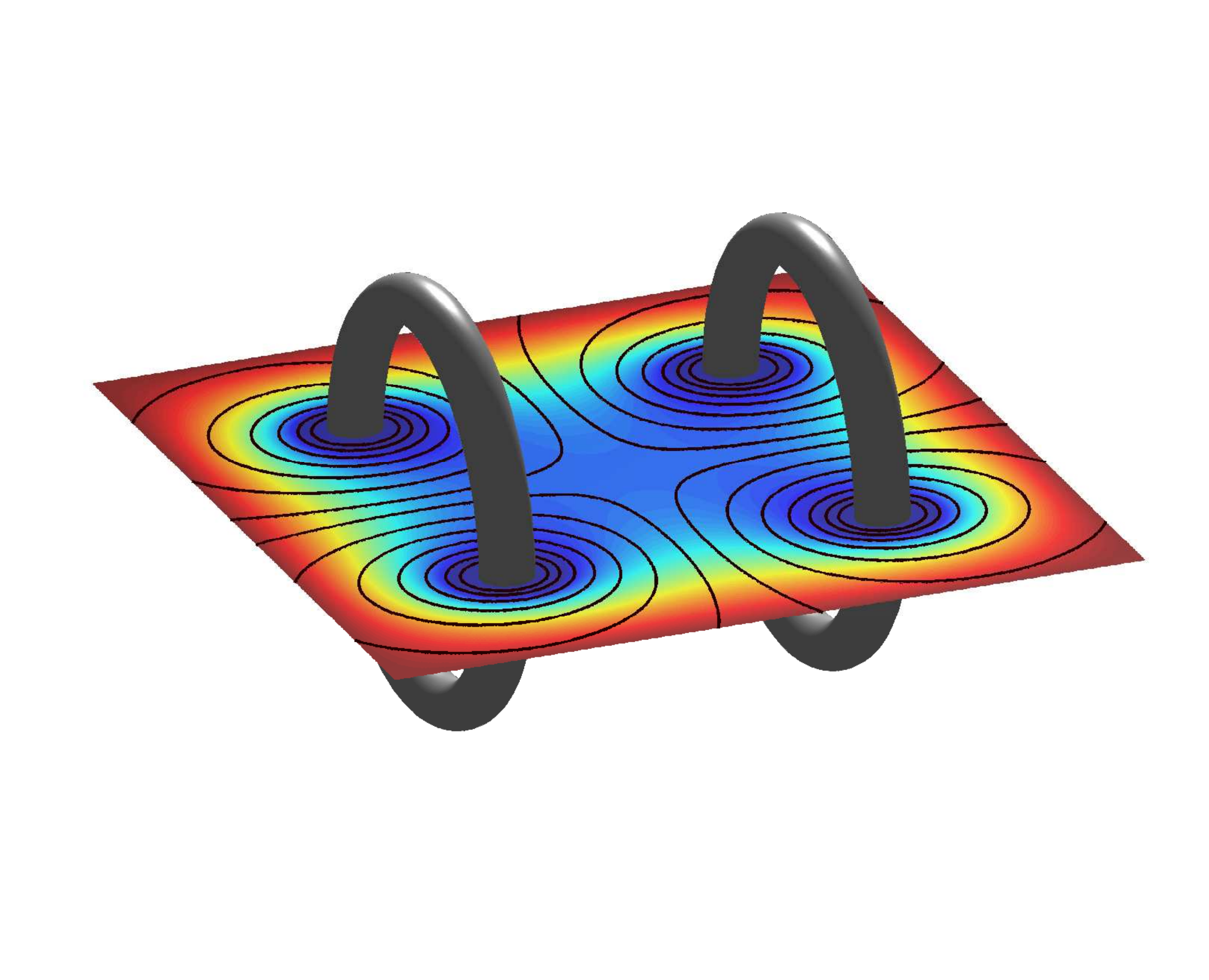}
\vspace{-15mm}
\caption{\label{fig2}Two-ring cathode grid with embedded magnetic
field coils that produce a magnetic cusp. Magnetic field lines
are in black; electric potential is shown as a heatmap with
blue representing a negative potential and red representing
a ground (a grounded cylindrical chamber is implicit at the edges of the
diagram)}
\end{figure}

We will examine the simple grid arrangement consisting of two
coaxial negatively biased rings within a grounded cylindrical
vacuum chamber as shown in Fig.\ref{fig2}, which
has been shown~\cite{khachan1, khachan2, shrier1} to give IEC
plasma properties. This grid replaces the cathode shown in
Fig.\ref{fig1}.

Current-carrying coils are embedded within each ring and
electrically isolated from it. Current circulates in opposite
directions in each coil
so as to produce opposing magnetic fields. The magnetic field created by the
coils will be shown to shield the surface of the rings from ion
impact. Moreover, the opposing currents
create a magnetic null at the midpoint of the axis through the
rings, point cusps at the center of each ring, and a line cusp in
a plane that bisects the axis at right angles.

The purpose of this
magnetic configuration is to increase the ion density beyond the
limit imposed by a concentration of only positive charges.
Trapping electrons in this cusped geometry, such that their
density is lower than those of the ions, enables the ion density
to be increased while maintaining the net positive charge on the
non-neutral plasma. This increase in ion density will enable an
increase in fusion density. Although electrons are not needed in
order to show a net energy gain, they enable an increase in the
net energy produced for practical considerations.

In general, the device is similar to a spindle cusp arrangement
but with a negative voltage bias imposed on the rings. Although
other grid configurations can be used, the arrangement shown here
has cylindrical symmetry that makes analysis and construction
relatively straightforward.

In the following subsections we will quantify the various energy
gain and loss mechanisms, then finally integrate all of them to
illustrate a net fusion energy gain.

\subsection{Density profiles and fusion power}
The conserved quantities (constants of the motion) for our system
are the total energy, $H$, and the $z-$~component of angular
momentum, $L_z$, which  correspond to time- and
rotation-invariance, respectively. In cylindrical polar
coordinates $(r,\theta, z)$ with $z$ the axis of symmetry, we
have:
\begin{align}
H &= \frac{1}{2}m(v_r^2 + v_z^2 + v_\theta^2) + q \phi \\
L_z &= r ( mv_\theta + qA_\theta ),
\end{align}
where $\phi$ is the electric potential and $A_\theta$ is the only
nonvanishing component of the magnetic vector potential for our
system. Density variation emerges through the position-dependence
of $A_\theta$ and $\phi$. $L_z$ is seen to suppress density around
the $r=0$ axis.

Where the magnetic field is slowly-changing over distance
scales of order of the gyroradius, we have a
third (``adiabatic'') invariant: $\mu \approx \frac{mv_\perp^2}{2B}$, $v_\perp$ is the component of particle velocity perpendicular to the local magnetic field.
Adiabaticity will tend to reduce cusp losses~\cite{kaye1}; we will assume for the sake of conservatism that our plasma is nonadiabatic.

Since any function whose only arguments are constants of the motion is
a stationary solution to the Vlasov equation, we can
write electron and ion distribution functions in the form $f = f(H, L_z)$
if we neglect individual collisons and treat only collective behaviour.
The collisional effects will later be estimated at the steady state of
the collisonless system.

For this paper we consider pseudo-Maxwellian distributions of the form
\begin{equation}
\begin{aligned}
f(H, L_z) =&  n_0\sqrt{\frac{2m}{2\pi^3 kT}} \exp\left\{-H/kT\right\} \exp \left\{ - L_z^2 / (2 m r_0^2 kT )\right\} \\
&\times \operatorname{erfc}\left(\beta(H - H_c)/kT\right)
\label{eq:distribution}
\end{aligned}
\end{equation}

Here $r_0$ represents a characteristic radius at which angular momentum becomes significant; one might expect this to be of the order of the particle gyroradius. Spatial dependence is also implied through $\phi$ and $A_\theta$ and thus $H$. This form for the distribution, though very simple, will suffice to demonstrate the parameter-dependence of our system. The gradient and value of the cutoff are determined by $\beta$ and $H_c$ respectively. Beyond this cutoff, particles are assumed to be able to escape from confinement. These pseudo-Maxwellian distributions can be expected to yield an underestimate for achievable fusion rate.

Recognising the symmetry between $v_r$ and $v_z$ in Eq.~\ref{eq:distribution}, we can rewrite it as
\begin{equation}
\begin{aligned}
f_\vec{v}  (v_L, v_\theta, A_\theta, \phi, r) = &\frac{2n_0}{\sqrt{\pi}} \sqrt{\frac{m}{2 k T}} v_L \\
& \times \exp\left\{{-\frac{m}{2kT}(v_L^2 + v_\theta^2 +  \frac{2 q\phi}{m})}\right\} \\
& \times  \exp\left\{{-\frac{1}{2 m kT} (r/r_0)^2(mv_\theta + qA_\theta)^2}\right\}  \\
& \times \operatorname{erfc}\left( \frac{\beta m}{2kT}(v_L^2 + v_\theta^2 +  \frac{2 q\phi}{m}) - \frac{\beta H_c}{kT} \right)
\end{aligned}
\end{equation}

with $v_L^2  = v_r^2 + v_z^2$. In this $(v_L, v_\theta)$ velocity-space, the angle-averaged velocity difference between two points $\vec{v_1}$ and $\vec{v_2}$ is
\begin{equation}
\langle \vec{v_1} - \vec{v_2} \rangle = \sqrt{ v_{1L}^2 + v_{2L}^2 }\hat{\vec{v}}_L + (v_{1\theta} - v_{2\theta})\hat{\vec{v}}_\theta
\label{eq:velocityDiff}
\end{equation}

Changing scales to remove dimensions, let:
\begin{displaymath}
\begin{aligned}
&v \to v' \equiv v / {\sqrt{2 kT/m}}\\
&\phi \to \phi' \equiv \frac{ e \phi }{kT} \\
&H_c \to H'_c \equiv \frac{ H_c }{kT} \\
&A_\theta \to A_\theta' \equiv e A_\theta / \sqrt{2 m k T} \\
&r \to r' \equiv r / r_0
\end{aligned}
\end{displaymath}

Dropping the primes (we will now work in the rescaled quantities):
\begin{equation}
\begin{aligned}
f_\vec{v}(v_L, v_\theta, A_\theta, \phi, r) =&\,\frac{2n_0}{\sqrt{\pi}}v_L \exp\left\{-(v_L^2 + v_\theta^2 + Z\phi)\right\}  \\
& \times \exp\left\{-r^2 (v_\theta + Z A_\theta)^2\right\}  \\
& \times  \operatorname{erfc}\left(\beta\left\{v_L^2 + v_\theta^2 + Z\phi - H_c \right\}\right)
\end{aligned}
\end{equation}

with $Z$ the atomic number ($1$ for a deuteron, $-1$ for an electron).

In the limit where $H_c \to \infty$, the density is given by
\begin{equation}
\begin{aligned}
n(A_\theta, \phi, r) &= \int_{-\infty}^{\infty} dv_\theta\, \int_{0}^{\infty} dv_L\,  \lim_{H_c \to \infty} f_\vec{v}(v_L, v_\theta, A_\theta, \phi, r) \\
&= \frac{n_0}{\sqrt{r^2+1}} \exp\left\{-\left(Z^2A_\theta^2 \frac{r^2}{r^2+1} + Z\phi \right)\right\}
\end{aligned}
\end{equation}

The presence of the cutoff complicates this somewhat; after some algebra and taking the limit $\beta \gg 1$, we arrive at
the approximate form, derived in appendix A:
\begin{equation}
\begin{aligned}
\lim_{\beta \to \infty} &n(A_\theta, \phi, r) = \frac{\sqrt{\pi}}{2} \exp\left\{-\left(r Z A_\theta\right)^2\right\} \left\{ \Bigg. \right.\\
&\frac{1}{\tilde{r}} \exp\left\{(r^4/\tilde{r}^2)Z^2 A_\theta^2 - Z\phi\right\}\, \mathcal{C}(\frac{r^2}{\tilde{r}} Z A_\theta, \tilde{r}w) \\
& - \frac{1}{r} \exp\left\{r^2 Z^2 A_\theta^2 - H_c\right\}\, \mathcal{C}(rZA_\theta, rw) \left.   \Bigg. \right\} \\
&\text{where}\ \, w = \Re\left( \sqrt{H_c - Z\phi} \right) \ge 0,\ \  \tilde{r} = \sqrt{r^2 + 1} \\
&\text{and}\ \, \mathcal{C}(a,b) = \operatorname{erf}(a + b) - \operatorname{erf}(a - b)
\end{aligned}
\end{equation}

Imaginary values for $w$ are forbidden as they would represent plasma with potential energy beyond the energetic cutoff at $H_c$.

A sufficient condition for magnetic shielding is $\mathcal{C} \ll 1$ near the coil surface, i.e:
\begin{equation}
Z A_\theta > \left(w + \frac{1}{r}\right)
\label{eq:shieldingCondition}
\end{equation}

. Eq.~\ref{eq:shieldingCondition} is more readily satisfied as the device becomes larger.

Restoring dimensions, we can evaluate Eq.~\ref{eq:shieldingCondition} along the plane of a coil at half the coil radius, using the approximation for a current loop of radius $a$~\cite{jackson1}:
\begin{equation}
A_\theta(r = a / 2, z=0) \approx \left(\frac{4}{29}\right) \mu_0 I
\end{equation}

After a small amount of algebra, Eq.~\ref{eq:shieldingCondition} reduces to the following (in S.I.):
\begin{equation}
I > \frac{29 \sqrt{2m} \left( \sqrt{H_c - q\phi_{\text{coil}}} + 2\sqrt{kT} \left(\frac{r_0}{r_{\text{coil}}}\right) \right)}{4 \mu_0 q}
\label{eq:shieldingConditionExplicit}
\end{equation}

For a system of two opposed coils spaced approximately a radius apart, the current required differs from this by only a few percent. It is worth noting the scaling with the
square root of energy.

Given the ion distribution $f_i$, fusion power is given by the standard expression
\begin{equation}
P_{\text{fus}} = \frac{1}{2} E_{\text{fus}} \int d^3 x \ n_i(x)^2 \langle \sigma v \rangle\ ,
\end{equation}
where $E_{\text{fus}}$ is the fusion energy and
\begin{equation}
\begin{aligned}
&\langle \sigma v \rangle =  (1 / n_i(x)^2) \times \\
 \iint d \vec{v} \iint d \vec{v'} &\|\langle \vec{v} - \vec{v'}\rangle\|\sigma(\|\langle \vec{v} - \vec{v'}\rangle\|) f_i(\vec{v}) f_i(\vec{v'}).
\end{aligned}
\end{equation}

To qauntify the above we adopt the fit to the D-D fusion cross-section due to Duane~\cite{duane1}.

\begin{figure}[h]
\includegraphics[width=\columnwidth]{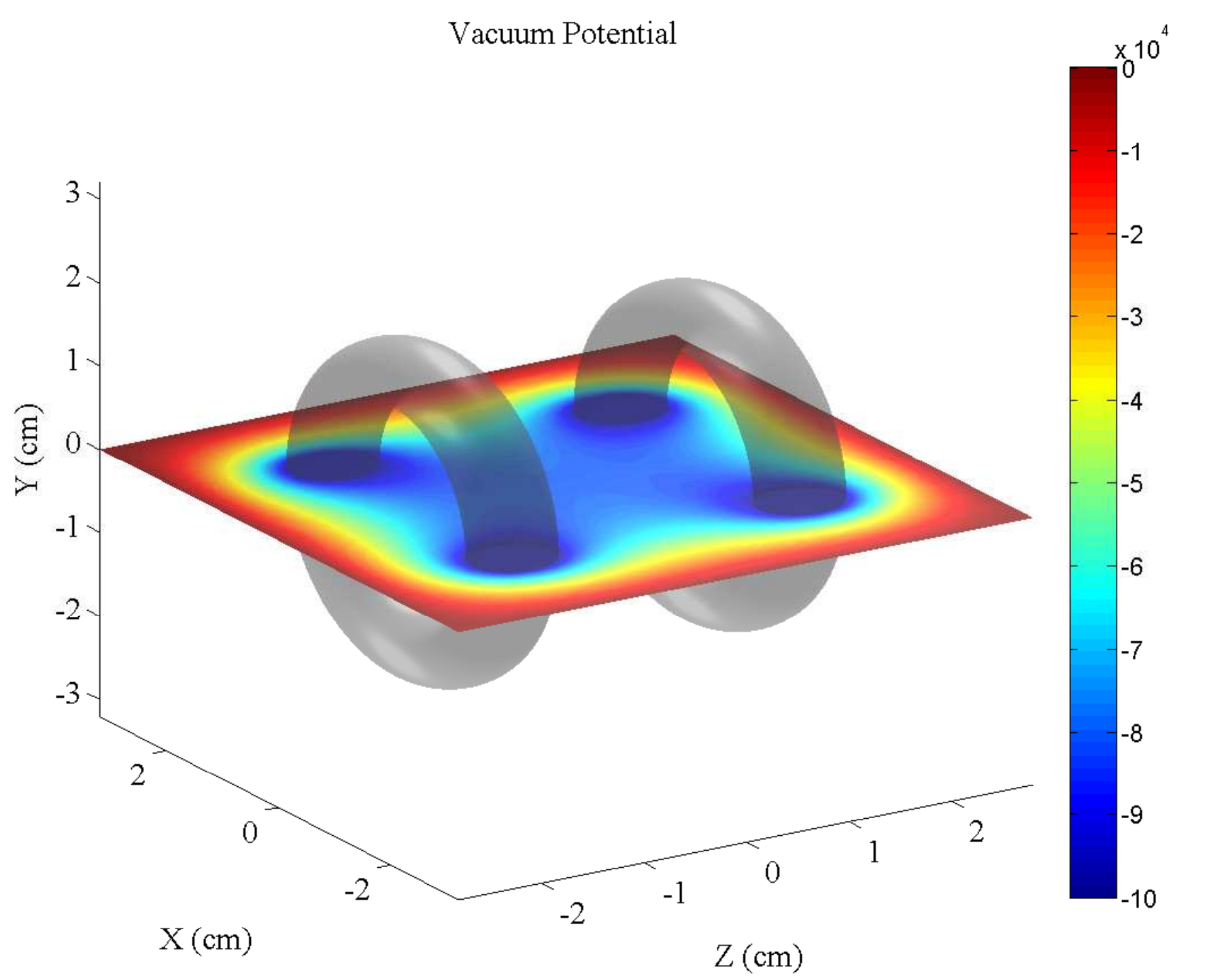}
\caption{\label{vacuumpotential}
Example Vacuum electric potential of charged cusp.
}
\end{figure}

\begin{figure}[h]
\includegraphics[width=\columnwidth]{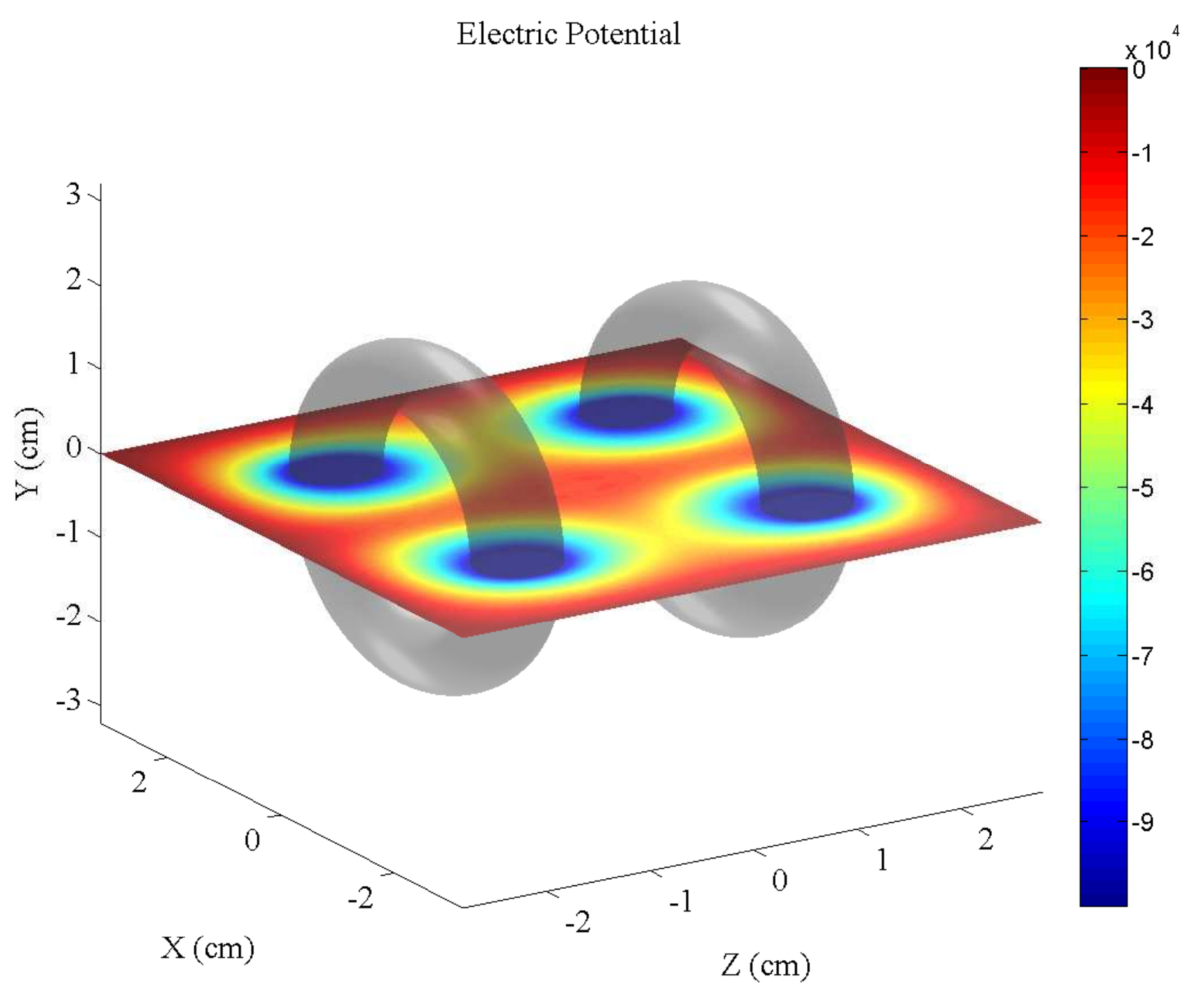}
\caption{\label{plasma}
Electric potential of example system with plasma present.
}
\end{figure}

\subsection{Cusp Losses}
Since typical operation of an electromagnetic cusp device involves the formation of a central potential, we need to obtain the
appropriate expressions for cusp confinement where an electric field is also present. For clarity, we work in S.I. in this section.

Derivation of the loss-cone angle begins with the force on a diamagnetic particle:
\begin{equation}
{\bf F_\parallel} = -\mu \nabla B - q \nabla \phi
\end{equation}

Since charge is conserved, energy conservation yields the separate conservation of the magnetic moment $\mu$.

We consider the motion of a particle turning around between fields $B_0, \phi_0$ and $B', \phi'$.
Let
\begin{eqnarray}
{\bf v_0} &=&{{\bf v_{\perp}}}_0 + {\bf v_{\parallel}}_0 \\
{\bf v'} &=&{\bf v'_{\perp}}
\end{eqnarray}

where ${\bf v_0}$ and ${\bf v'}$, respectively, are the velocities
at $B_0$ and $B'$.

Energy and magnetic moment conservation, respectively, yield
\begin{eqnarray}
\frac{1}{2}m {v'_{\perp}}^2 + q\phi' &=& \frac{1}{2}mv_0^2 + q\phi_0 \\
\frac{{{v_{\perp}}}_0^2}{B_0} &=& \frac{{{v'_{\perp}}^2} }{B'}
\end{eqnarray}

Combining and rearranging, we get
\begin{equation}
\frac{B_0}{B'} = \frac{{{v_{\perp}}}_0^2}{v_0^2 + \frac{2q}{m}(\phi_0 - \phi')}
\end{equation}

which yields the corrected loss-cone angle as
\begin{equation}
\sin^2\theta_m = \frac{B_0}{B_m}\left(1 + \frac{2q}{mv_0^2}(\phi_0 - \phi_m)\right) \equiv \frac{1}{{R_m}}
\label{eq:losscone}
\end{equation}
where $\theta$ is the angle between the velocity vector ${\bf v_0}$ and the
magnetic field in the low-field region, and $B_m$ is the maximum magnetic
field strength encountered by the particle during its motion.
If we can hold the coils
at a negative potential with respect to the centre of the device,
the combined electromagnetic trapping is sufficient to eliminate
electron cusp losses for electrons of energy below a small
multiple of the virtual anode height. Depending on the geometry,
such a virtual anode can form naturally when the coils are held at
a negative potential with respect to the boundary (e.g vacuum
chamber).

\subsection{Grid Losses}

Since the magnetic field of a cusp is convex inward, both the
curvature and $\nabla B$ drifts are nowhere towards the coils.
We can prevent a particle from reaching the coils by insisting
that the gyroradius becomes smaller than the distance of approach.

We recall the definition of the gyroradius
\begin{equation}
r_g = \frac{m v_\perp}{q B}
\end{equation}

Approximating the magnetic field near the current-loops with that due to a line-segment of current, we have
\begin{equation}
B(s) = \frac{\mu_0 I}{2 \pi s}
\end{equation}

with $s$ the perpendicular distance from the line-segment.

We require $r_g(r) < r - r_c$ for some $r_c$ representing the thickness (minor radius) of the coils; solving for I yields:
\begin{equation}
I > \frac{2 \pi m v_\perp}{\mu_0 q (1 - \frac{r_c}{r})} \approx \frac{2 \pi \sqrt{2 E_i m_i}}{\mu_0 q (1 - \frac{r_c}{r})}
\label{eq:currentRequired}
\end{equation}

Inspecting Eq.~\ref{eq:currentRequired}, we see that the required number of Ampere-turns depends on the achievable
separation between metal surfaces. This suggests the use of coils with a small minor radius.

If we take $r' = 2 r_c$ with $r'$ the distance at which our condition applies, then the above simplifies to
\begin{equation}
I > \frac{4 \pi \sqrt{2 E_i m_i}}{\mu_0 q_i}.
\label{eq:requiredCurrentSimplified}
\end{equation}

We can rearrange Eq.~\ref{eq:requiredCurrentSimplified} to find a critical energy for a given current, below which particles cannot escape to the grid:
\begin{equation}
E_{\text{crit}} = \frac{\mu_0^2 q_i^2 I^2}{32 \pi^2 m_i}
\end{equation}

The structure of this agrees with our earlier approach through the distribution function (Eq.~\ref{eq:shieldingConditionExplicit}).

If $E_{\text{crit}} > - q_i \phi_{\text{coil}}$, upscattered ions will be lost to the chamber walls before they are lost to the grid.

\subsection{Bremsstrahlung}
Bremsstrahlung is a significant loss mechanism for IEC systems; both efficiency and safety demand that our system does not radiate all the input energy as X-rays.

For electron-electron bremsstrahlung we
can use the result of Gould~\cite{gould1, gould2}:
\begin{equation}
\frac{P_{brem,\rm {e,e}}}{V} = \frac{20(44 -
3\pi^2)}{9\sqrt{\pi}}\alpha^3 (\frac{\hbar}{mc})^2 \sqrt{m_e}
n_e^2 {(kT_e)}^{3/2}
\end{equation}
with $\alpha$ the fine structure constant, $T_e$ the electron temperature.

In the case where an ion species is in thermal equilibrium with the electrons, an expression for electron-ion bremmstrahlung power density, due once again to Gould is
\begin{equation}
\frac{P_{brem,\rm {e,i,eq}}}{V} =
\frac{32}{3}\sqrt{\frac{2}{\pi}}\alpha^3 \frac{\hbar^2 c}{m} n_e
Z_i^2 n_i \tau(1 + \frac{11 \tau}{24})
\end{equation}
with $\tau = kT_e / mc^2$.

In our device, the ions and electrons are not necessarily in thermal equilibrium. Where the ions are slower than the electrons,
the effect on bremsstrahlung is small. Where the ions are faster than the electrons we can simply replace $\tau$ in the above with $\tau' =
(T_i m_e / T_e m_i ) \tau$\, i.e using the approximate electron
temperature in the rest-frame of the ion as suggested by Jones~\cite{jones1}. The electron-ion Bremsstrahlung is computed using the larger of $\tau$ and $\tau'$;
this selection is performed at each point in space.

\subsection{Collisional up-scattering}

Interspecies energy transfer and collisionally-induced upscattering both represent energy-loss mechanisms for an IEC-style device. We can obtain useful estimates of the magnitudes of such losses from the Landau equation for the scattering of particles of distribution $f_a$ due to distribution $f_b$~\cite{zhdanov1}:
\begin{equation}
\begin{aligned}
&\frac{\partial f_{a}(\vec{v})}{\partial t} = \frac{c_{ab}}{m_a} \nabla_{\vec{v}} \cdot \vec{J}(f_a, f_b, \vec{v}) \\
&\text{with} \\
&\vec{J}(f_a, f_b, \vec{v}) = \\
&\int \int d v'_L\, d v'_\theta  \left( \frac{\mathcal{I}}{u} - \frac{\vec{u}\vec{u}}{u^3} \right) \cdot \left[ \frac{f_b(\vec{v'})}{m_a} \nabla_{\vec{v}}f_a - \frac{f_a(\vec{v})}{m_b} \nabla_{\vec{v'}}f_b  \right]
\end{aligned}
\label{eq:LandauEquation}
\end{equation}

where $\mathcal{I}$ denotes the unit dyadic, $\vec{u}\vec{u}$ is the tensor product of $\vec{u}$ with itself, and
\begin{equation}
c_{ab} = \frac{Z_a^2 Z_b^2 e^4 \ln(\Lambda)}{8 \pi \epsilon_0^2}
\end{equation}
\begin{equation}
\ln(\Lambda) = \ln(4 \pi \epsilon_0 kT \frac{m_e}{m_a} \langle {v_a}^2 \rangle \lambda_D / q_e^2 )
\end{equation}
\begin{equation}
\vec{u} = \langle \vec{v} - \vec{v'} \rangle \ \ \text{(c.f Eq.\ref{eq:velocityDiff})}.
\end{equation}

For our distribution, the derivatives are as follows:
\begin{equation}
\frac{\partial f}{\partial v_L} = \left[\frac{1}{v_L} - 2 v_L \left( 1 + K(H) \right) \right] f
\equiv k_L f
\end{equation}

and

\begin{equation}
\frac{\partial f}{\partial v_\theta} = -2\left[ \Big. v_\theta\left(1 + r^2 + K(H) \right) + r^2 Z A_\theta \right] f \equiv k_\theta f
\end{equation}

where
\begin{equation}
K(H) = \frac{{2\beta} \exp\left\{-\beta^2(H - H_c)^2\right\}}{ {\sqrt{\pi}}\operatorname{erfc}\left(\beta \left(H- H_c\right)\right)  }
\end{equation}
with $H = v_L^2 + v_\theta^2 + Z\phi$. The term $K(H)$ is a pure consequence of our cutoff. Without such a cutoff, it is seen
that upscattering and heating power may be dramatically underestimated.

Recalling that the average energy of a particle of species $a$ is by definition
\begin{equation}
\langle E_a \rangle = \frac{1}{n} \int f(\vec{v}) E(\vec{v})\ d^3\vec{v}
\end{equation}

the total heating power experienced by species $a$ is
\begin{align}
P_{\text{heat, a}}(x) &= n_a(x)\frac{\partial \langle E_a \rangle}{\partial t} = \int \int d v_L\, d v_\theta \frac{\partial f_a(\vec{v})}{\partial t} E(\vec{v})  \\
&=  -c_{ab} \int_{-\infty}^{\infty} d v_\theta \int_0^{\infty} d v_L\, \vec{v} \cdot \vec{J}(f_a, f_b, \vec{v})
\label{eq:pheat}
\end{align}

Here, use has been made of the product rule for continuously differentiable functions:
\begin{equation}
\int_{\Omega} \frac{\partial u}{\partial x_i} v \,d\Omega = \int_{\Gamma} u v \, \hat\nu_i \,d\Gamma - \int_{\Omega} u \frac{\partial v}{\partial x_i} \, d\Omega,
\end{equation}

where $\Gamma$ is the boundary of volume $\Omega$. In each of our particular cases at least one of these terms vanishes.

Up-scattering-driven loss power can be computed in the same fashion with merely a change of integration limits such that we integrate over that region of the distribution holding particles that can escape and replace $E(\vec{v})$ with a bound on the kinetic energy at the chamber wall.
\begin{equation}
\begin{aligned}
&P_{\text{upscattering loss for species a}} \\
&= \sum_b {c_{ab}} \int_{\infty}^{\infty} d v_\theta \times \\
& \left(v_\theta^2 + H_c\right) \hat{\vec{v}}_L \cdot \vec{J}\left(f_a, f_b, \vec{v} = \left(w_a, v_\theta\right)\right)
\label{eq:upscattering}
\end{aligned}
\end{equation}

Note that we remain in dimensionless units; to get power in Watts, we multiply the above by $kT$ (expressed in Joules).

\subsection{Power balance}

Each fusion event consumes two ions. The number of fusion events per unit time is $R_{\text{fus}} = P_{\text{fus}} / E_{\text{fus}}$. Maintaining energy balance requires we replace fused ions with fresh ions at the pre-collision energy. Since this is necessarily less than the maximum ion kinetic energy, we can bound this by:
\begin{equation}
P_{\text{ion loss}} < P_{\text{ion upscattering loss}} + 2 P_{\text{fus}} \frac{ | q_i \phi_{\text{coil}} |}{E_{\text{fus}}}
\end{equation}

Let $Q \equiv \frac{\text{Useful Power Out}}{{\text{Power in}}}$.

If the fraction of fusion power which can be captured is $\eta_f$ then
\begin{equation}
\text{Useful Power Out} = \eta_f P_\text{fusion}
\end{equation}

Power loss in the coils will be omitted under the assumption that a real fusion
power system will employ superconducting magnets. We will treat only the case
where ions exert a net heating effect on the electrons; we will then add the electron heating
power to the electron loss power.
The continuous power input required by a fusion system such as has been described is at most
\begin{equation}
\text{Power in} = P_{\text{brem}} + P_{\text{electron heating}} + P_{\text{ion loss}} + P_{\text{electron loss}}
\end{equation}

Thus
\begin{equation}
Q = \frac{\eta_f P_\text{fusion}}{P_{\text{brem}} + P_{\text{electron heating}} + P_{\text{ion loss}} + P_{\text{electron loss}}}
\end{equation}

The combined-cycle efficiency of a modern gas-turbine approaches $60\%$ - we will thus take $\eta = 0.6$ as a representative choice.

\section*{Results}

For this paper, we restrict ourselves to low densities; in particular, we ensure that the particle currents are small enough that we can use the vacuum values for the magnetic vector potential. Where this is true, we can solve iteratively for a self-consistent density and electrostatic potential structure by repeatedly recomputing densities and averaging the resulting $\phi$. The electrostatic potential for a given density profile, coil potential, and grounded chamber is itself computed through an over-relaxation developed for this paper. When computing electron densities, the value $H_c$ for the electron population has been continuously adjusted such that $Z\phi >= H_c$ everywhere on the interior of the device and $Z\phi = H_c$ at at least one point on the boundary.

Electrons outside the device will be both fast-moving and rapidly lost to the walls. To model this, electron densities in the region exterior to the coils have been held at zero during the relaxation.

Once the potentials $A_\theta$ and $\phi$ have been obtained everywhere on our spatial lattice, numerically integrate is performed to compute the gain and loss terms introduced in Eqns.~\ref{eq:pheat} and \ref{eq:upscattering}. In the process, we derive spatial maps of the various quantities.

\begin{figure}[h]
  \includegraphics[width=\columnwidth]{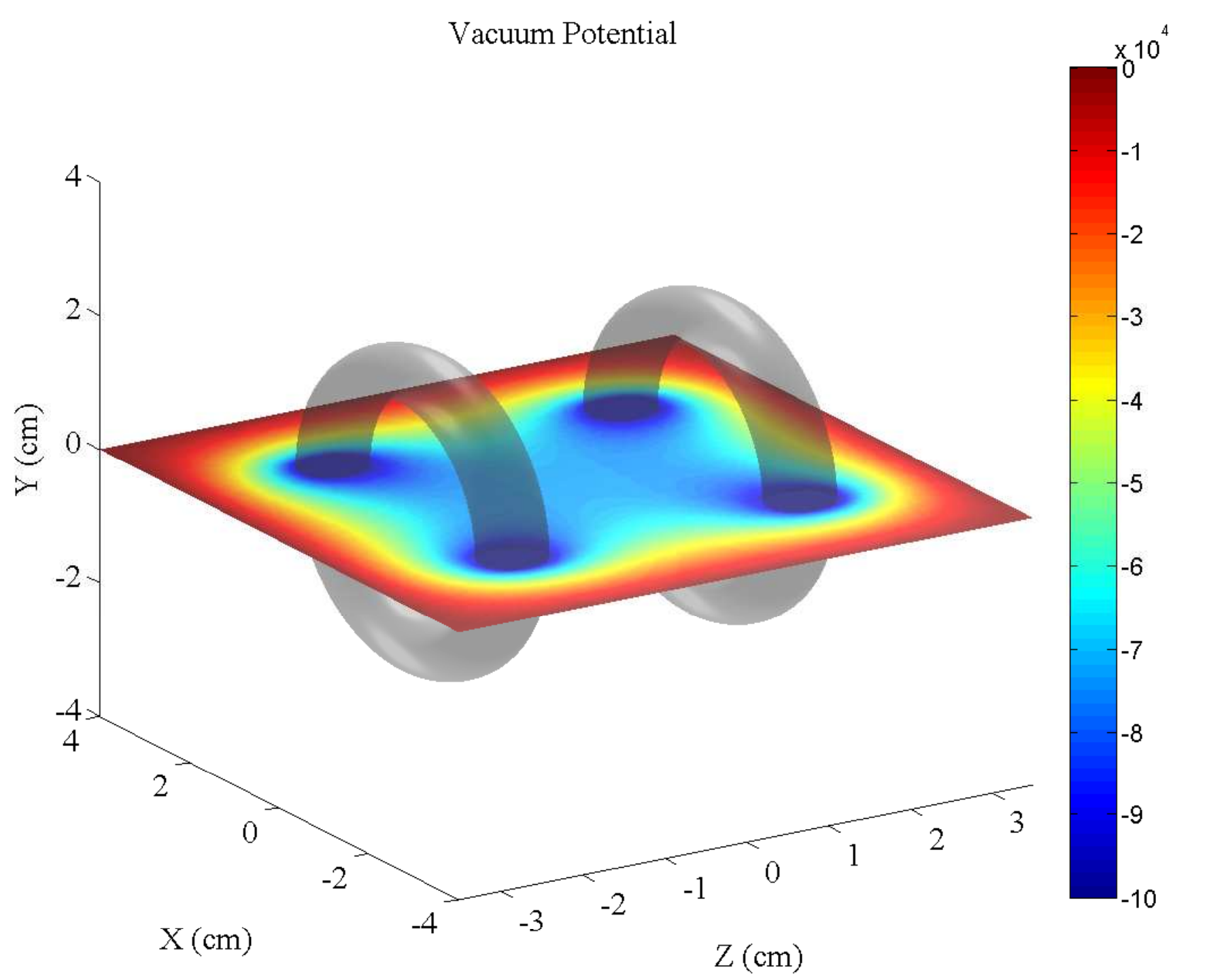}
  \caption{\label{fig:3D_VirtualCathodePhi_Vacuum}Vacuum electric potential of electron-free virtual-cathode configuration}
\end{figure}

\begin{figure}[h]
  \includegraphics[width=\columnwidth]{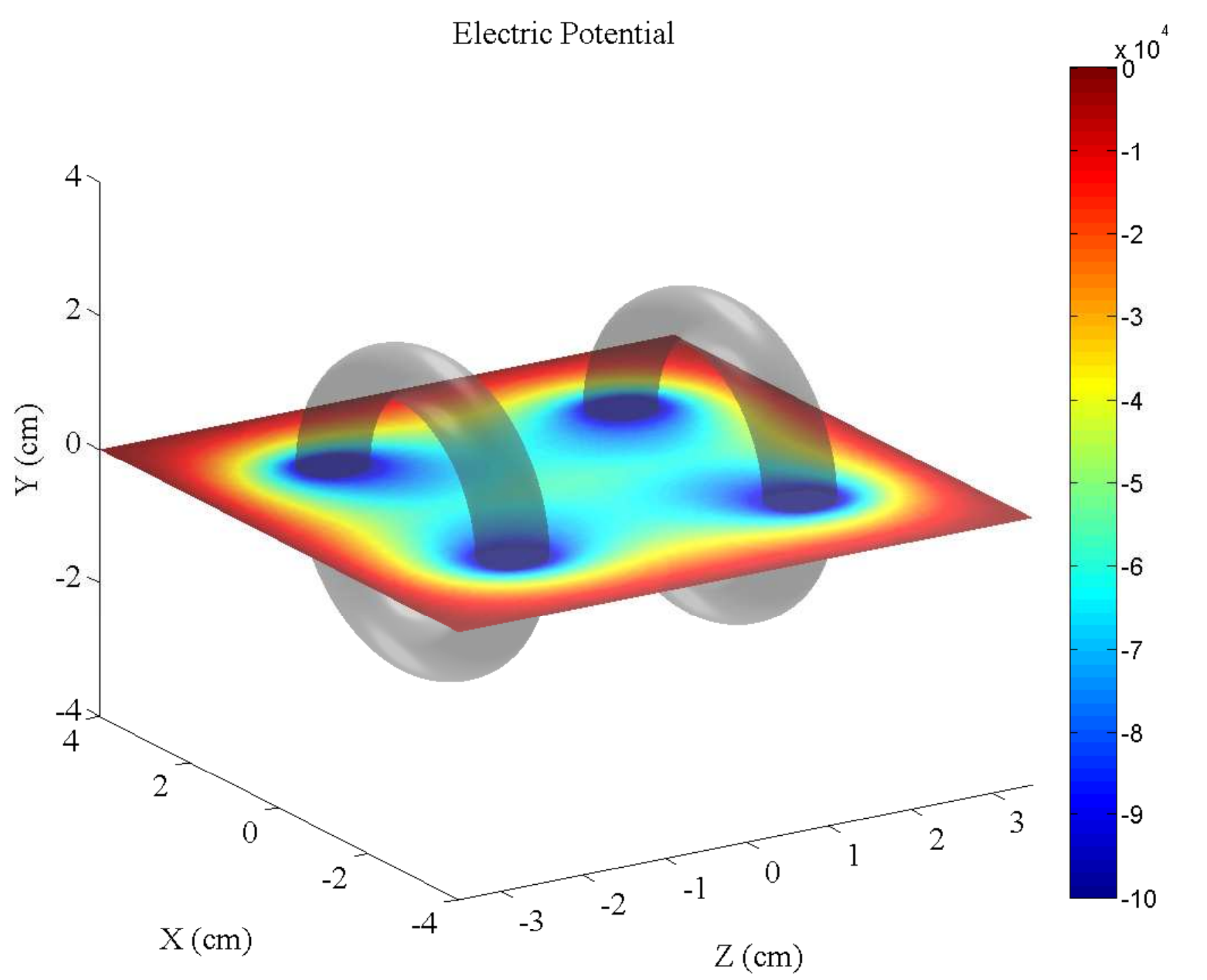}
  \caption{\label{fig:3D_VirtualCathodePhi}Electric potential of virtual-cathode configuration with ions present}
\end{figure}

\begin{figure}[h]
  \includegraphics[width=\columnwidth]{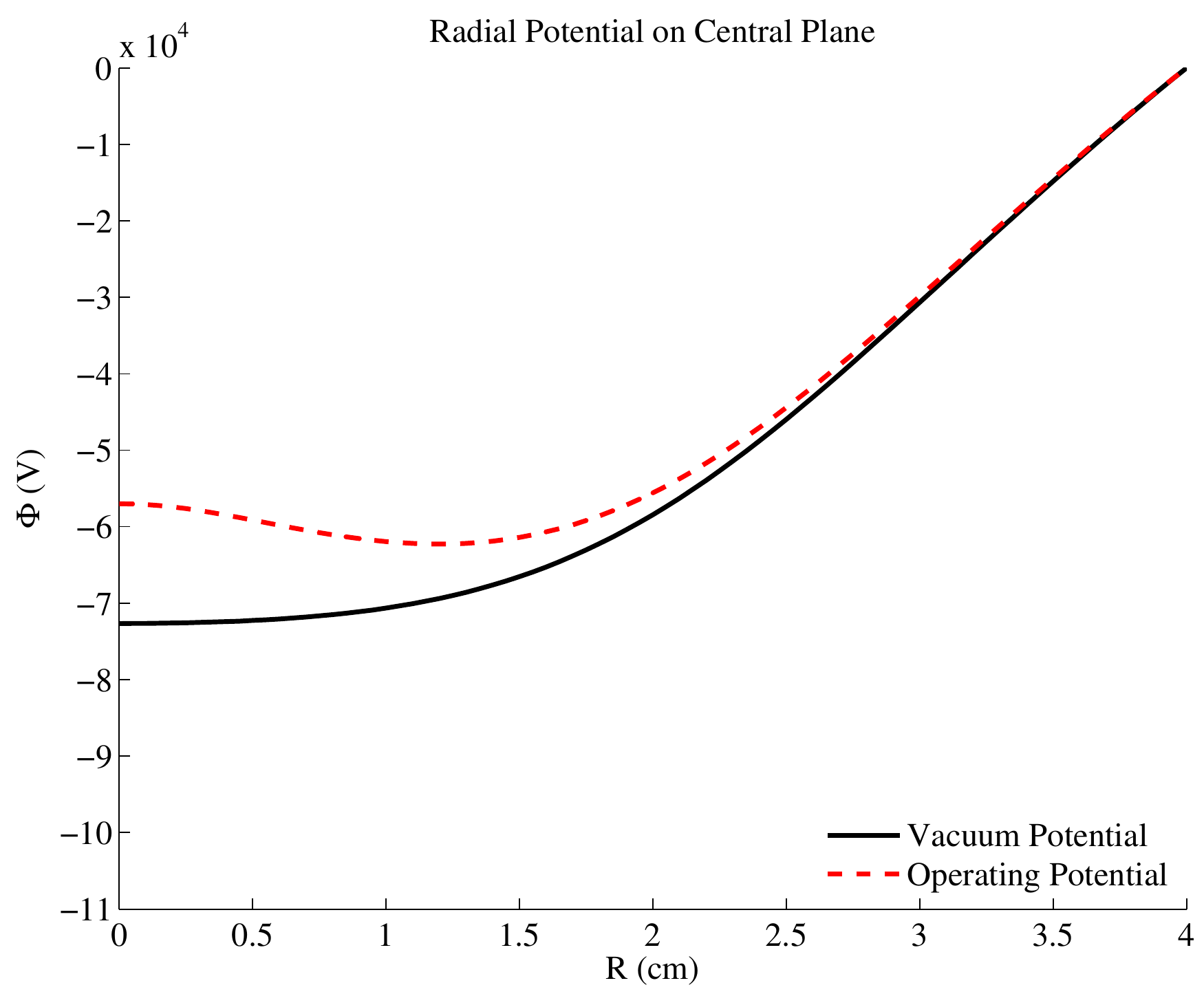}
  \caption{\label{fig:VirtualCathodeRSlice_Phi} Electric potential along a radial line through the midpoint of the device for electron-free configuration. }
\end{figure}

\begin{figure}[h]
  \includegraphics[width=\columnwidth]{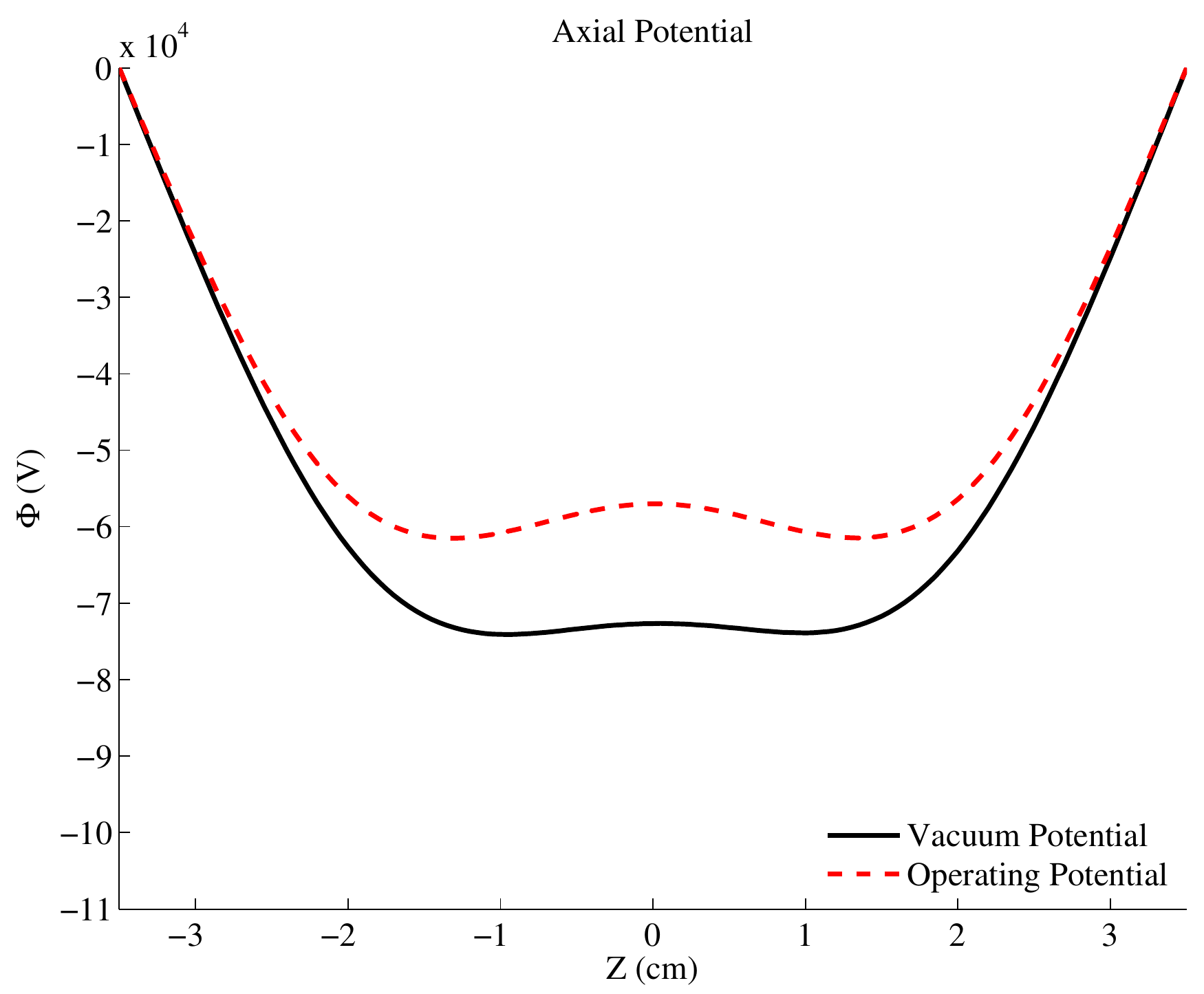}
  \caption{\label{fig:VirtualCathodeZSlice_Phi} Electric potential along an axial line through centre the device for electron-free configuration. }
\end{figure}

The first configuration examined is an electron-free system exhibiting a virtual cathode in the centre of the device. Parameters for this device are in table~\ref{table:virtualCathodeParams}. The potential-well structure with and without plasma is shown in Figs. ~\ref{fig:3D_VirtualCathodePhi_Vacuum}, ~\ref{fig:3D_VirtualCathodePhi}, ~\ref{fig:VirtualCathodeRSlice_Phi}, and ~\ref{fig:VirtualCathodeZSlice_Phi}. A local virtual anode is formed, but the potential of this local anode is lower than any point on the boundary.

\begin{figure}[h]
  \includegraphics[width=\columnwidth]{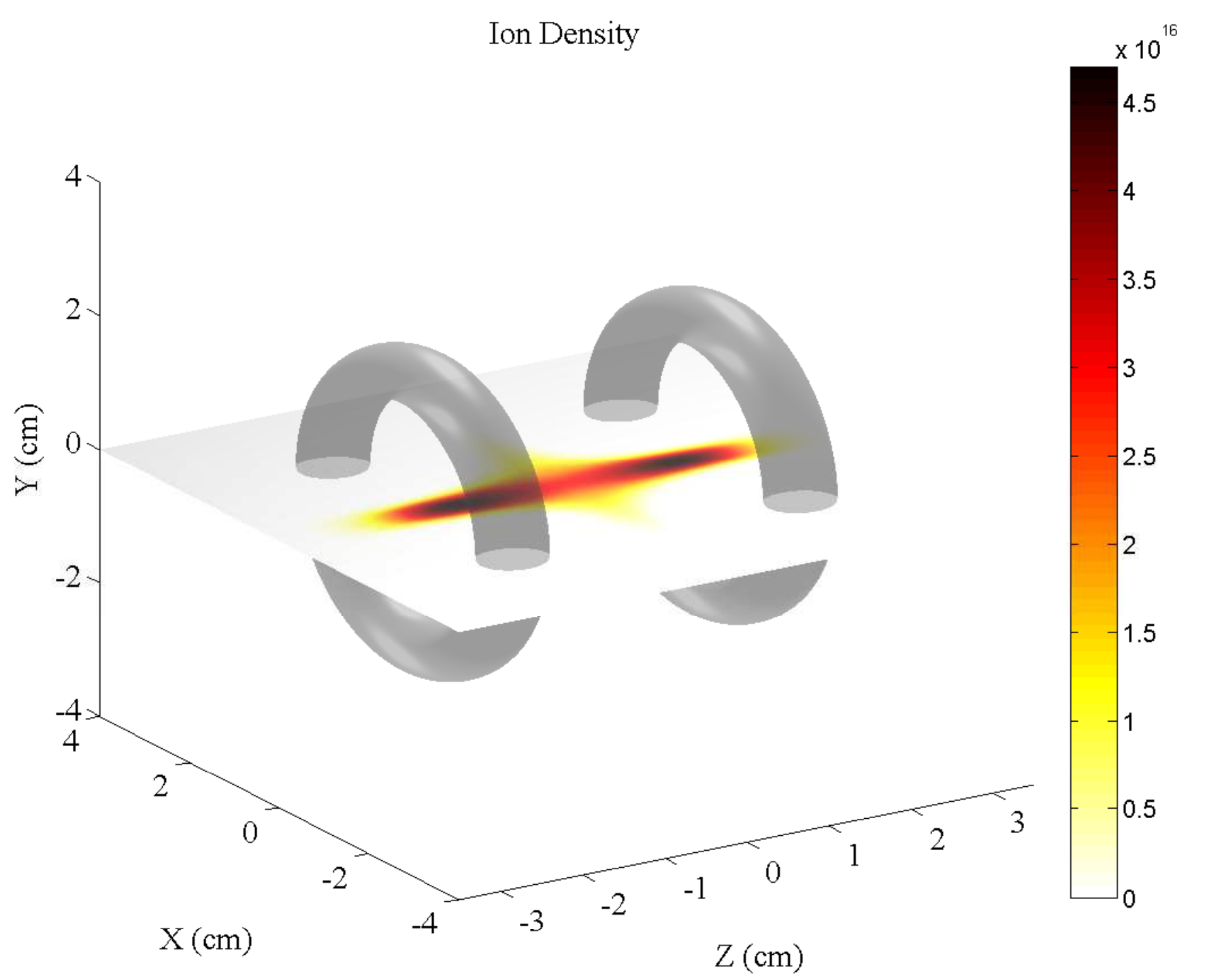}
  \caption{\label{fig:3D_VirtualCathodeIonDensity}3D slice of ion density for the  electron-free virtual-cathode configuration}
\end{figure}

\begin{figure}[h]
  \includegraphics[width=\columnwidth]{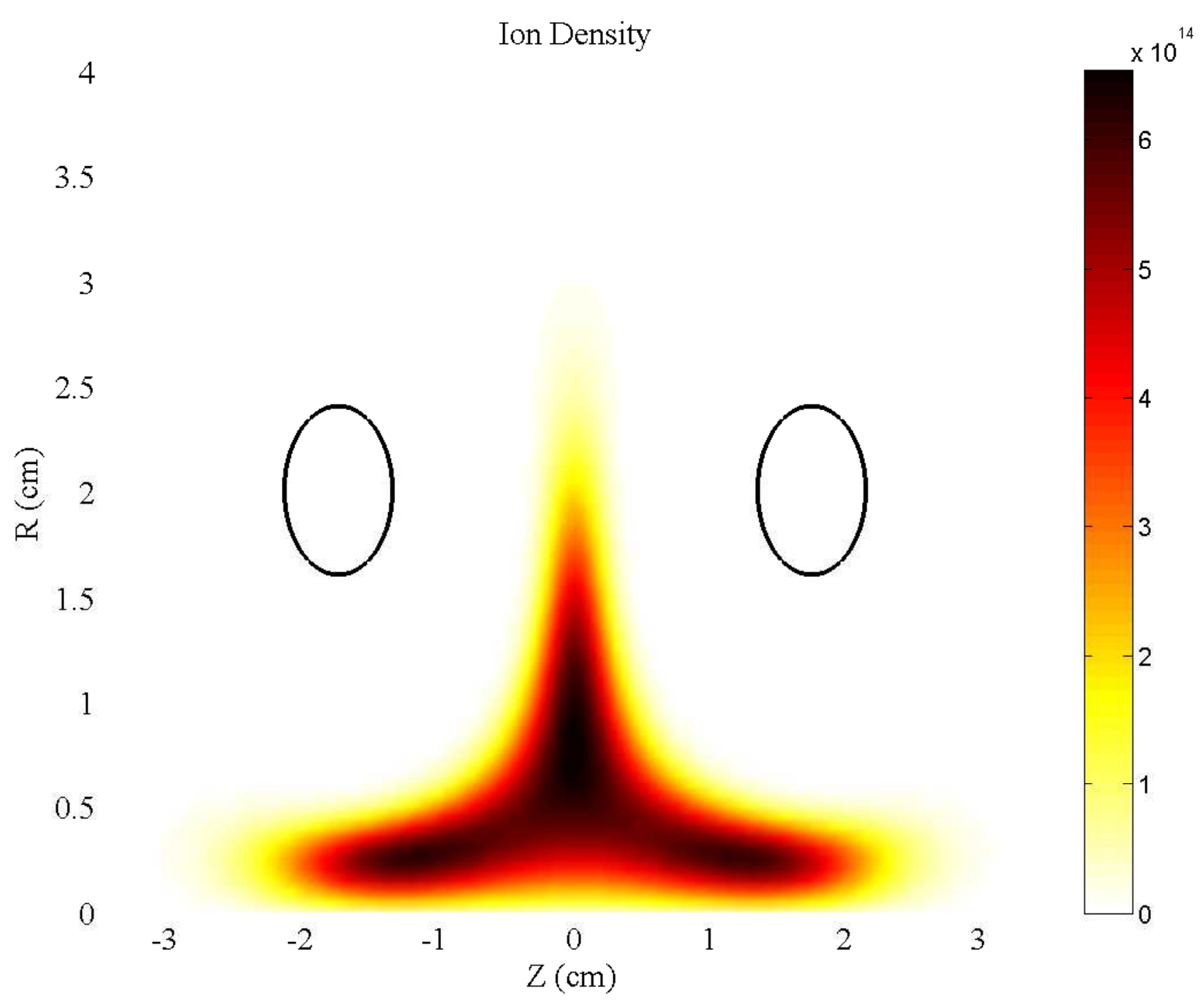}
  \caption{\label{fig:2D_VirtualCathodeIonDensity}Ion density in integrated (r,z) space for electron-free virtual-cathode configuration. Black circles show the coil cross-section.}
\end{figure}

The corresponding ion density is shown in Fig.~\ref{fig:3D_VirtualCathodeIonDensity}. It must be understood that these 3D pictures show a slice; the corresponding density integrated over the azimuthal angle is given in Fig.~\ref{fig:2D_VirtualCathodeIonDensity} showing the total contribution from different radii.

\begin{figure}[h]
  \includegraphics[width=\columnwidth]{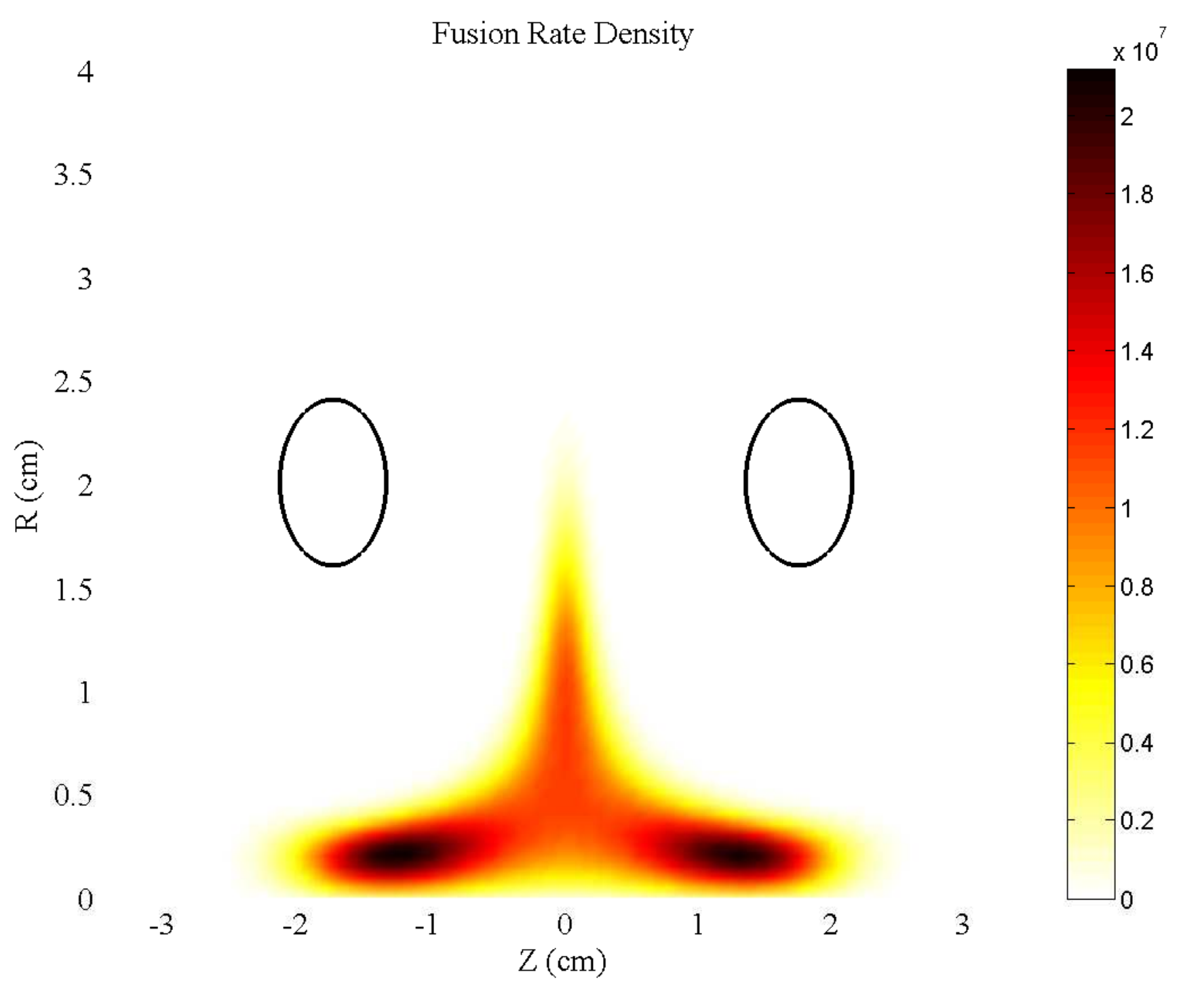}
  \caption{\label{fig:2D_VirtualCathodeFusionDensity}Fusion-rate density in integrated (r,z) space for electron-free virtual-cathode configuration. Black circles show the coil cross-section; most of the fusion is occuring on the axis near the planes of the coils.}
\end{figure}

Azimuthally-integrated fusion density for this configuration is depicted in Fig.\ref{fig:2D_VirtualCathodeFusionDensity}. Most of the fusion occurs close to the axis and near the planes of the coils. With no electrons to contribute to Bremsstrahlung, the dominant energy loss mechanism is the upscattering of ions; this is shown to be insignificant. The gain of the system reduces to $0.6 \times 2.5\, \text{MeV} / 100\, \text{keV} = 15$ as shown in Table~\ref{table:virtualCathodeParams}.

\begin{table}
\begin{tabular}{|l|l|}
\hline
Parameter & Value \\
\hline
Vacuum chamber radius & 40 mm \\
Vacuum chamber length & 70 mm \\
Coil major radius & 20 mm \\
Coil minor radius & 4 mm \\
Coil separation   & 35 mm \\
Coil current      & 700 kA-turns \\
Coil potential    & -100 kV \\
kT (ions)         & 10 keV \\
$r_0$ (ions)      & 5 mm \\
$\beta$ (ions)    & 100 \\
$H_c$ (ions)      & 0  \\
$E_\text{fus}$    & 2.5 MeV \\
spatial resolution & 0.5 mm \\
\hline
Computed Quantity & Value \\
\hline
Max. ion density  & $4.7 \times 10^{16}$ $\text{m}^{-3}$ \\
Total fusion rate  &  2650 / s \\
Total fusion power &  1.06 nW \\
Ion upscattering power & $1.0 \times 10^{-23}$ W   \\
Q & $\approx$ 15 \\
\hline
\end{tabular}
\caption{\label{table:virtualCathodeParams} Parameters for an electron-free system with a virtual cathode}
\end{table}

Introducing electrons allows higher densities through cancellation of the space-charge due to the ions. Parameters for an example system are given in Table~\ref{table:electronParams}. Bremsstrahlung power is by far the dominant loss mechanism in this system; one should expect the electrons in such a system to exhibit considerable cooling, reducing this loss. The proper treatment of this reduction is left to future work.

\begin{table}[h]
\begin{tabular}{|l|l|}
\hline
Parameter & Value \\
\hline
Vacuum chamber radius & 40 mm \\
Vacuum chamber length & 70 mm \\
Coil major radius & 16 mm \\
Coil minor radius & 4 mm \\
Coil separation   & 28 mm \\
Coil current      & 500 kA-turns \\
Coil potential    & -100 kV \\
kT (ions, electrons)    & 10 keV \\
$r_0$ (ions, electrons)      & 5 mm \\
$\beta$ (ions, electrons)    & 100 \\
$H_c$ (ions)      & 0  \\
$H_c$ (electrons)      & 20.9 keV  \\
$E_\text{fus}$    & 2.5 MeV \\
spatial resolution & 0.25 mm \\
\hline
Computed Quantity & Value \\
\hline
Max. ion density  & $1.86 \times 10^{17}$ $\text{m}^{-3}$ \\
Max. electron density  & $2.04 \times 10^{17}$ $\text{m}^{-3}$ \\
Total fusion rate  &  42360 / s \\
Total fusion power &  17 nW \\
Total Brem power &  0.4 nW \\
Electron upscattering power & $2.7 \times 10^{-4}$ nW \\
Electron heating power & $7 \times 10^{-4}$ nW \\
Ion upscattering power & $2.7 \times 10^{-9}$ nW \\
Ion upscattering power & $1.0 \times 10^{-23}$ W   \\
Q & $\approx$ 9 \\
\hline
\end{tabular}
\caption{\label{table:electronParams} Parameters for an electron/ion system with a virtual anode}
\end{table}

\begin{figure}[hp]
  \includegraphics[width=\columnwidth]{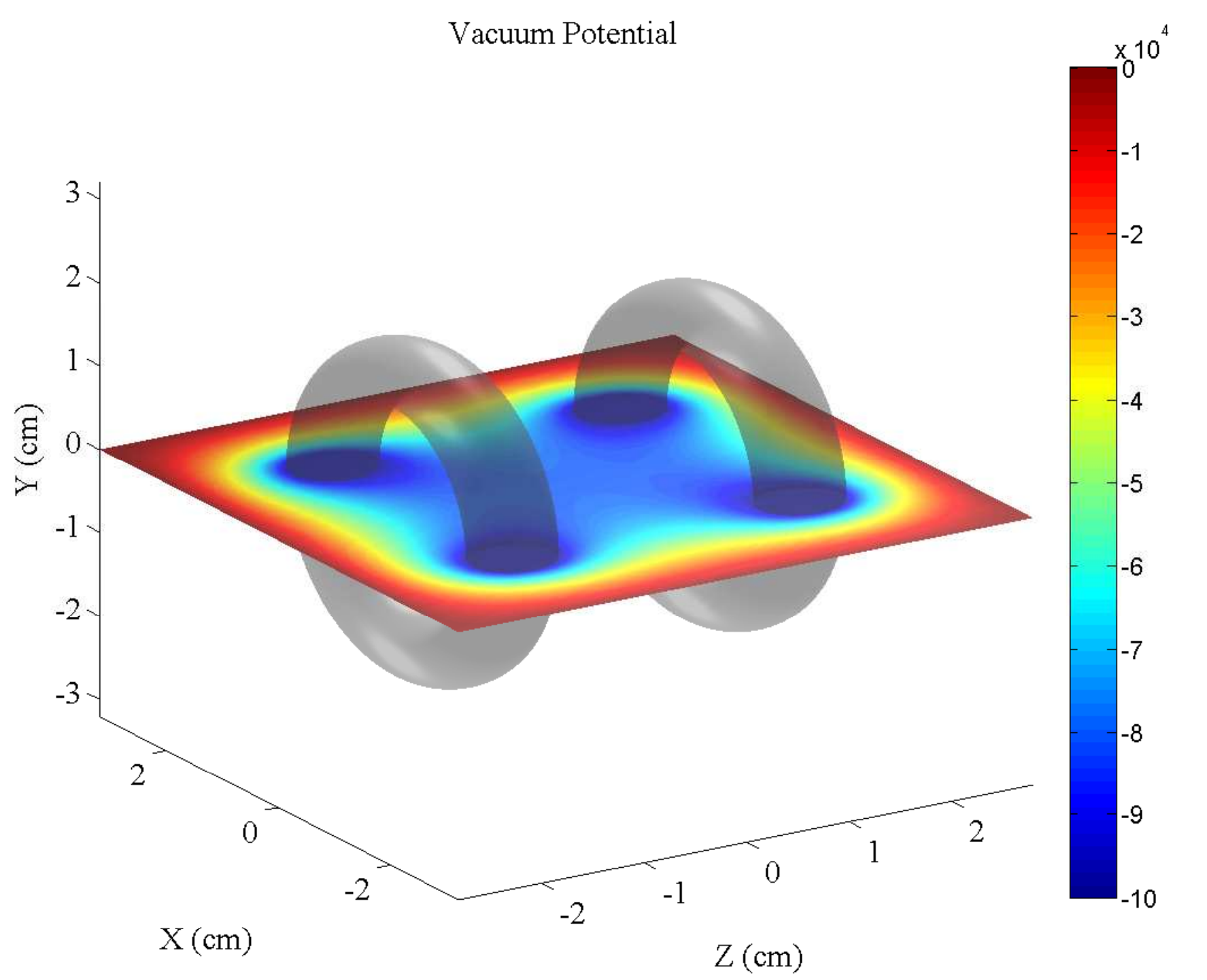}
  \caption{\label{fig:3D_phi_vacuum}Vacuum electric potential of electron/ion system}
\end{figure}

\begin{figure}[hp]
  \includegraphics[width=\columnwidth]{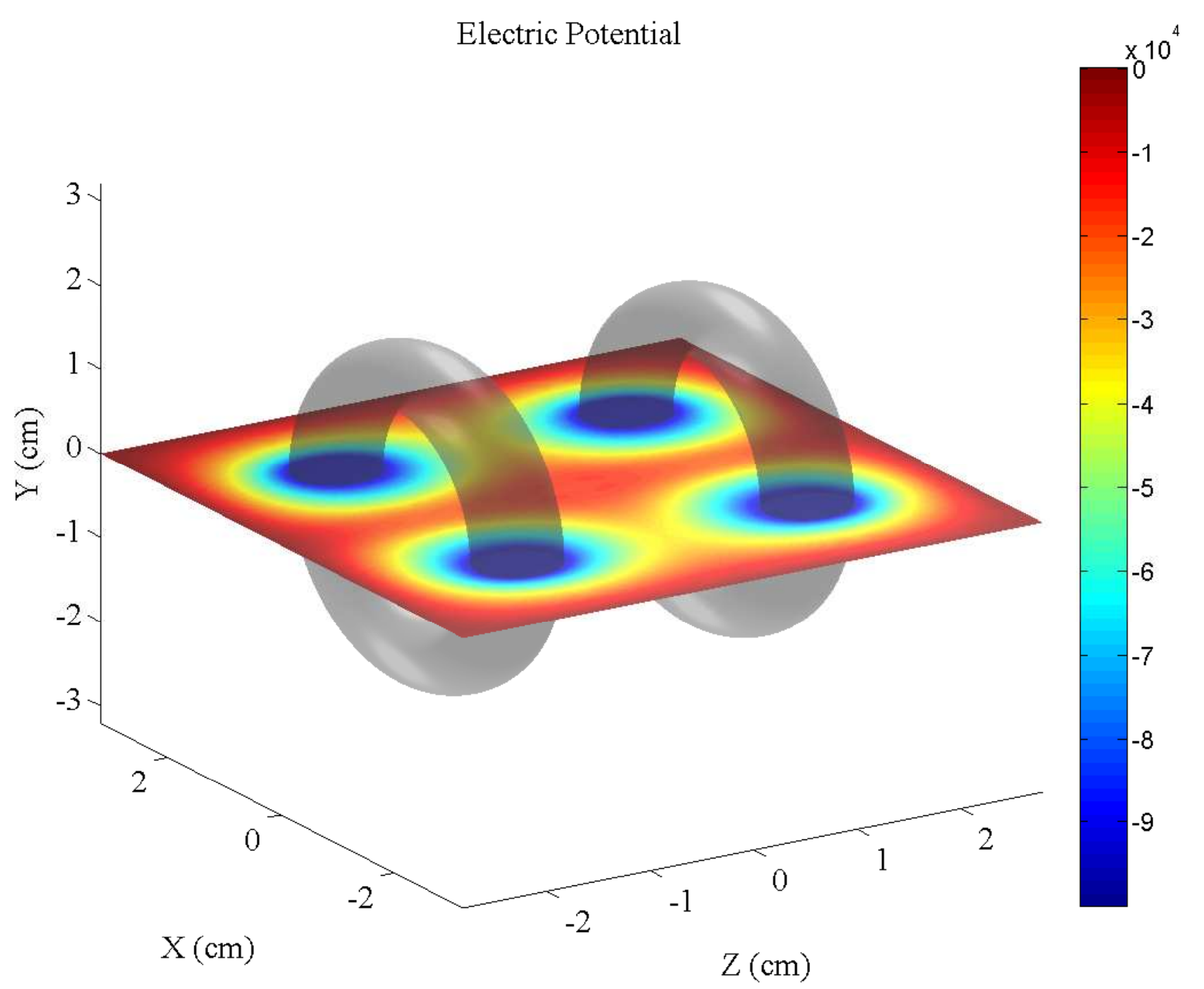}
  \caption{\label{fig:3D_phi}Electric potential of electron/ion system with plasma present}
\end{figure}

\begin{figure}[hp]
  \includegraphics[width=\columnwidth]{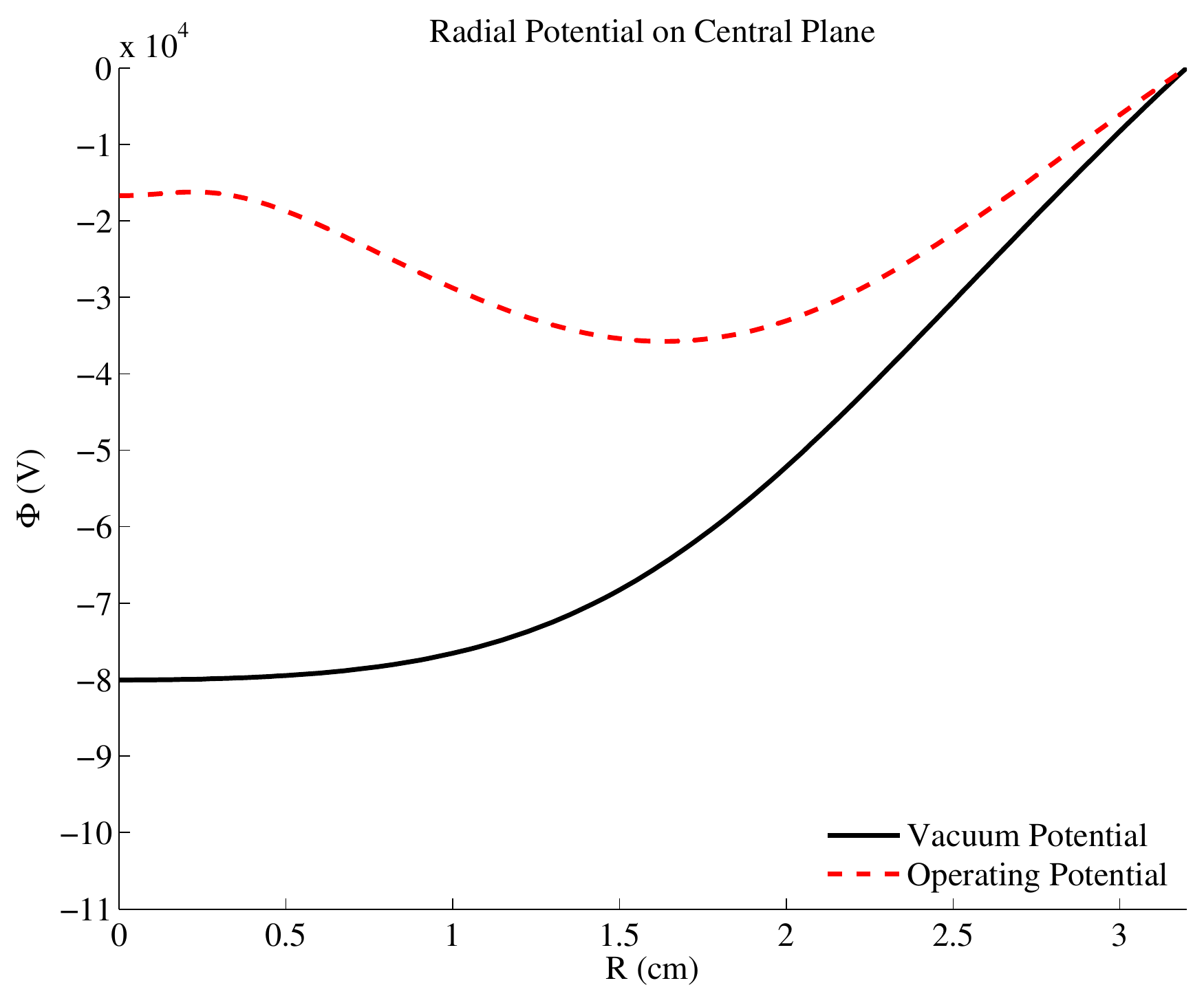}
  \caption{\label{fig:RSlice_Phi} Electric potential along a radial line through the midpoint of the device where electrons are present. }
\end{figure}

\begin{figure}[hp]
  \includegraphics[width=\columnwidth]{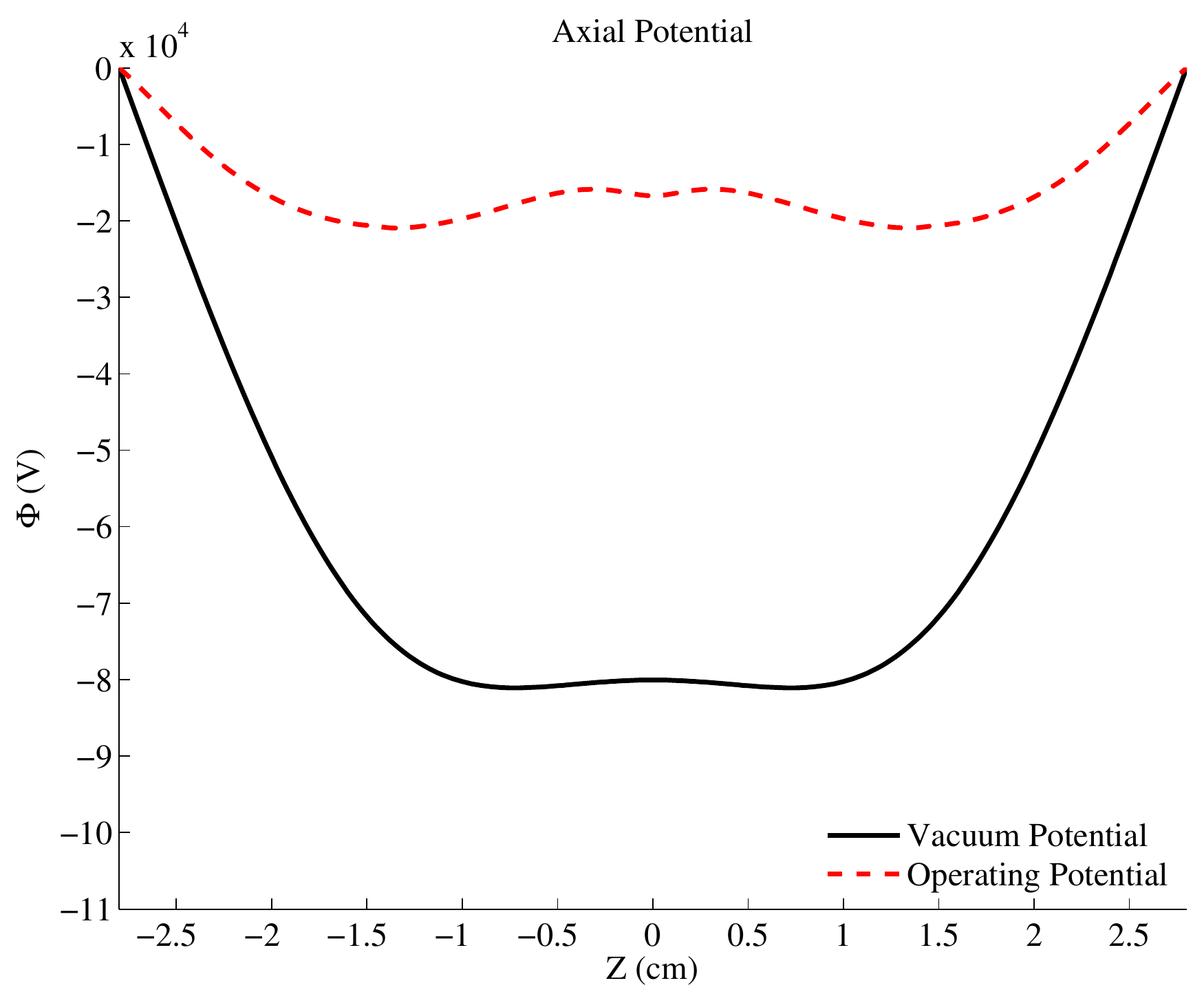}
  \caption{\label{fig:ZSlice_Phi} Electric potential along an axial line through centre the device where electrons are present. }
\end{figure}

\begin{figure}[hp]
  \includegraphics[width=\columnwidth]{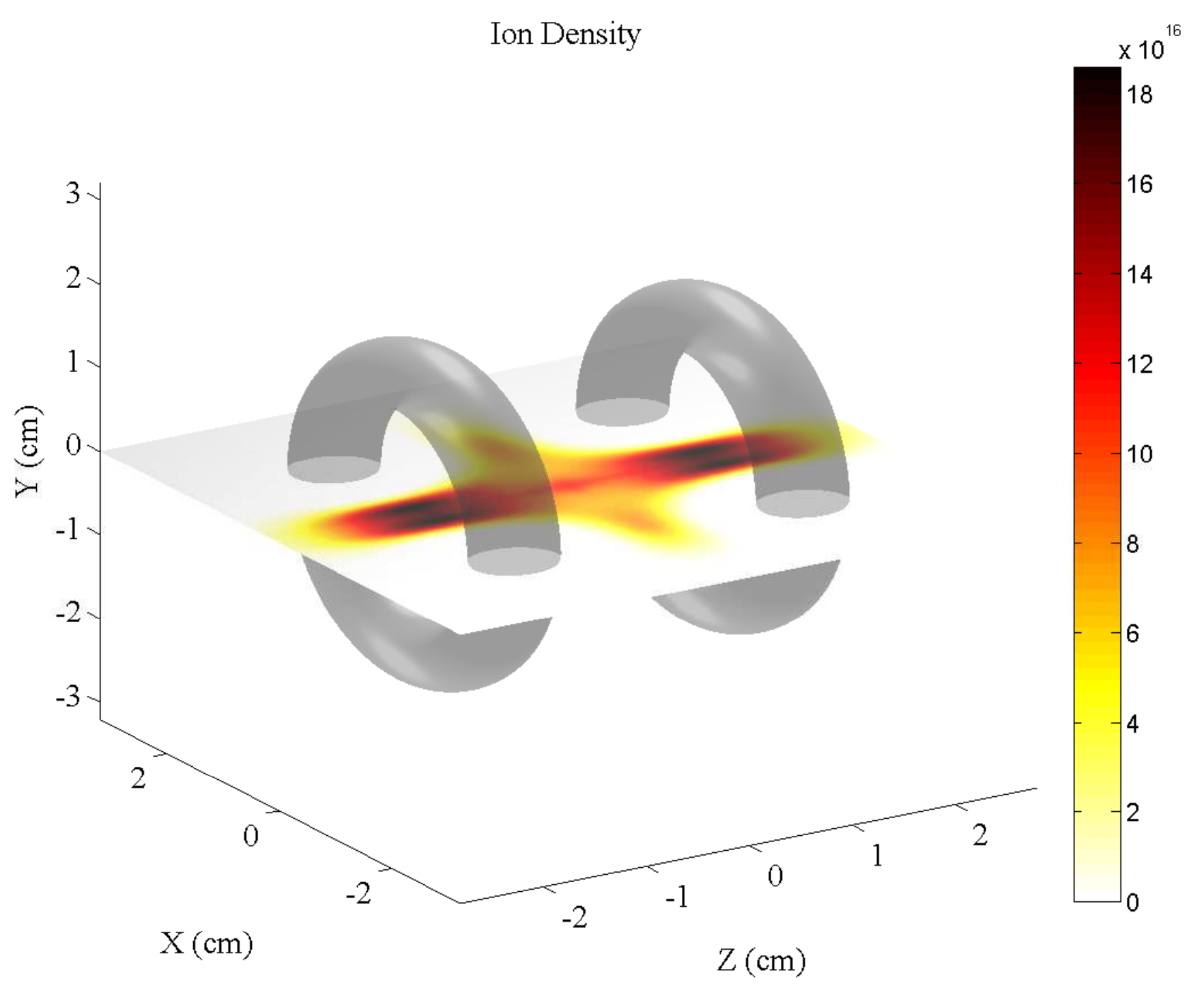}
  \caption{\label{fig:3D_ion_density} Ion density in the electron/ion system}
\end{figure}

\begin{figure}[hp]
  \includegraphics[width=\columnwidth]{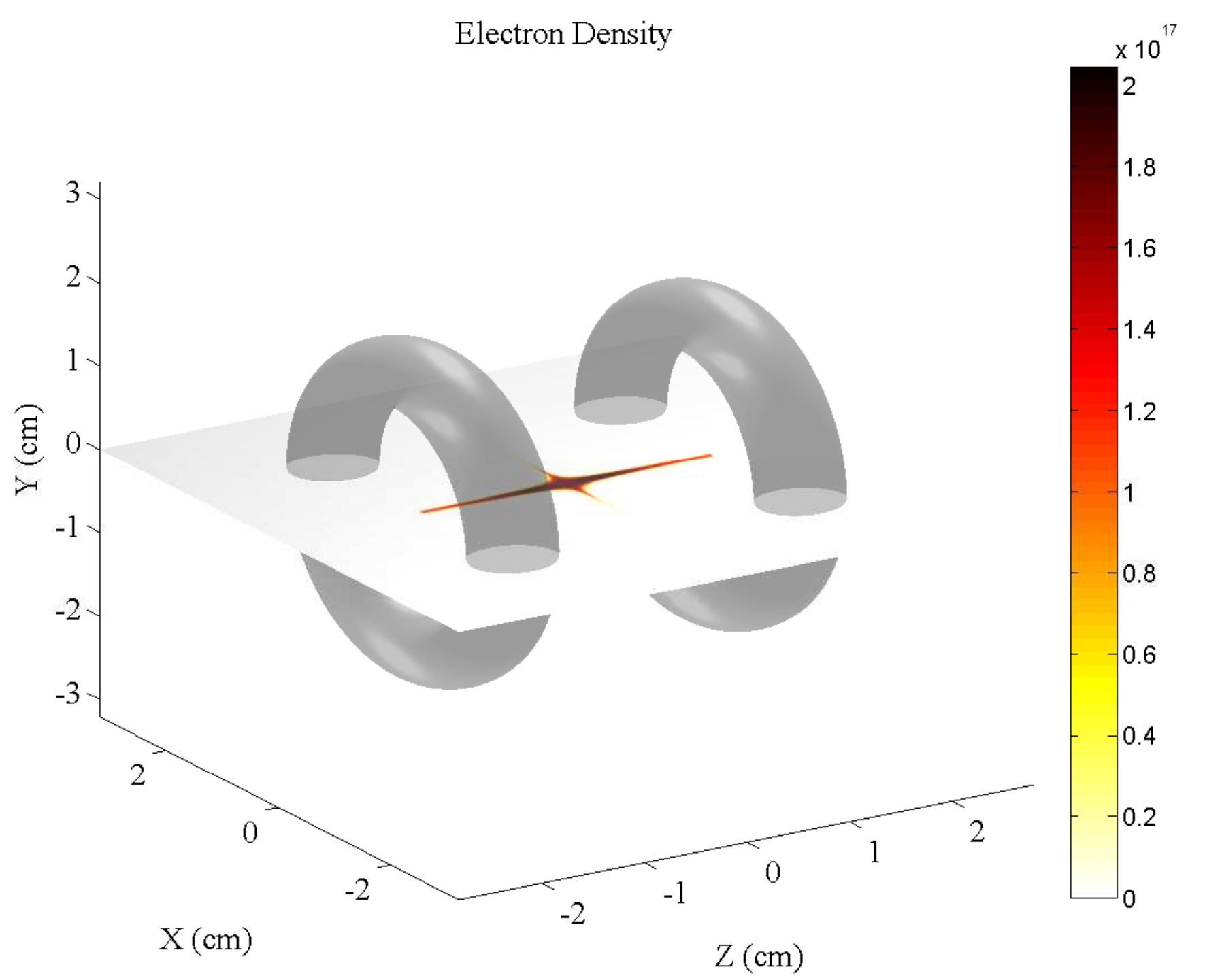}
  \caption{\label{fig:3D_electron_density} Electron density in the electron/ion system}
\end{figure}

The potential-well structure in Figs.~\ref{fig:RSlice_Phi} and \ref{fig:ZSlice_Phi} exhibits a small secondary virtual cathode - the ``Poissor'' structure proposed by Hirsch~\cite{hirsch1}.
The overall virtual anode is responsible for a great enhancement in electron confinement. Figs.~\ref{fig:3D_ion_density} and \ref{fig:3D_electron_density} show the
ion and electron densities, respectively. The formation of a jet-like structure is notable.

\begin{figure}[hp]
  \includegraphics[width=\columnwidth]{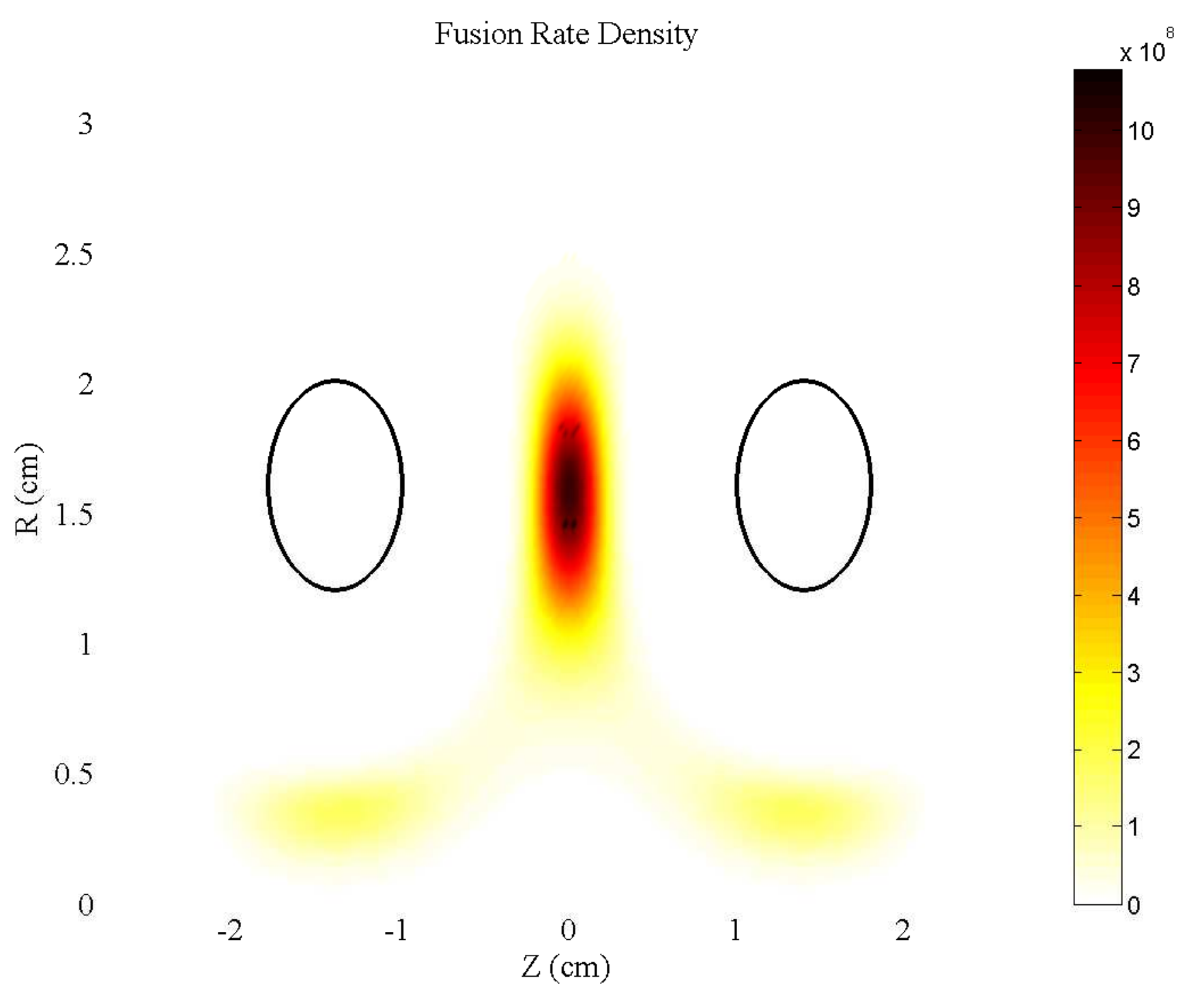}
  \caption{\label{fig:2D_fusion_rate_density} Fusion rate density in the electron/ion system}
\end{figure}

Fusion in this electron/ion configuration will occur in three rings, as show in Fig.~\ref{fig:2D_fusion_rate_density}. For a higher-density system,
we would alter the geometry to raise the potential in the ring cusp and consequently shift more of the ion density towards the center of the system.

In both the cases depicted the gain, $Q$ of the system is considerably greater than 1. No attempt has been made to optimise the parameters of the system; such a study awaits access to more computational power and extension of the density/potential solver to higher densities where simple fixed-point iteration is no longer applicable.

\section*{Conclusions and Outlook}

It has been shown that an IEC device can achieve energy gain if the electrodes are circular coils carrying sufficient current to prevent bombardment by and subsequent collection of charged particles. The required current can be estimated easily from the energy of the particles in the plasma. Gains of the order of $10$ are predicted for a deuterium-burning system at low density. The combination of electric and magnetic potentials in this case is sufficient to confine a significant electron population close to the device axis, admitting the possibility of a beamlike plasma with recirculating ions. The dimensions were chosen for numerical convenience - they should not be taken as representing limits to energy output. We see no reason why the device should not scale to MW outputs, with achievable fusion power bounded by thermal and mechanical properties of the materials used in construction.

Our results constrast with earlier work by Rider~\cite{rider1},
whose conclusions we regard as dubious, and which cannot be recovered unless we assume a Maxwellian, quasineutral plasma with
the density distribution due to Krall (derived for non-Maxwellian distribution) and neglect the use of the electrostatic potential for the mitigation of cusp losses. We are unable to imagine an IEC device consistent with all of Rider's assumptions; indeed concerns over self-consistency motivated much of our work.

Some caveats apply; we have deliberately ignored turbulence and instabilities; we have also ignored drifts which can arise through inhomogeneities in the plasma such as the diamagnetic drift. We quote results only for low densities of the order of mTorr and for power outputs in the nanoWatts. We have also assumed a form for the distribution which corresponds to a steady state of the Vlasov equation and thus also ignore the dynamics of the approach to equilibrium - the steady state of a driven system should not be expected to be identical to this.

In adopting the Landau integral form for collisional transport, we are emphasising grazing collisions over strong collisions. In practice, all fusion systems must exhibit strong, inelastic, collisions - otherwise they would produce no fusion. The proper treatment of collisions for our system should involve a deeper consideration of these scattering events; one approach might be a monte-Carlo simulation involving full electrodynamics.

Finally, we have assumed a fully-ionized system. A real fusion system will include neutral species; ionization and recombination will introduce a new loss mechanism. As very high levels of ionization can be achieved at the low densities explored in this paper, we leave the exploration of the neutral-species interactions for another paper.

Future work will investigate the transition from the low-$\beta$ regime to the high-beta cusp-confinement regime treated in other papers, and develop further the theory of plasma confinement in charged multipole cusps. It would be interesting to map the parameter-space boundary wherein $Q>1$. The modest volume and density requirements to produce measurable fusion rates suggest the cost of building an experimental device may not be prohibitive, either.

\section{Acknowledgements}

John Hedditch would like to thank Duncan and Annelies Hedditch for supporting him in this work, and David Gummersall for valuable discussion.

\newpage
\begin{widetext}
\appendix

Here we present the derivation of the density for our distribution away from the limit of infinitely-sharp cutoff.

\section{Density calculation}
Using the fact that
\begin{equation}
\int dx\, x e^{-x^2/c} \operatorname{erfc}(ax^2 + b) = - \frac{c}{2}e^{-\frac{x^2}{c}}
 \left\{e^{\left( \frac{x^2}{c} + \frac{b}{ac} + \frac{1}{4a^2c^2}\right)} \operatorname{erf}(ax^2 + b + \frac{1}{2ac}) + \operatorname{erfc}(ax^2 +b) \right\} + C
\end{equation}

we have
\begin{equation}
\int_0^{\infty} dx\, x e^{-x^2/c} \operatorname{erfc}(ax^2 + b) = \frac{c}{2}\left\{\operatorname{erfc}(b) - e^{ \left( \frac{b}{ac} + \frac{1}{4a^2c^2}  \right) } \operatorname{erfc}\left(b + \frac{1}{2ac}\right) \right\}
\end{equation}

and therefore

\begin{equation}
\begin{aligned}
\int_0^{\infty} dv_L\, F_L(v_L, v_\theta, \phi) &= \frac{1}{2} \left\{\operatorname{erfc}\left( \beta C(v_\theta, \phi) \right) - \exp\left\{{C(v_\theta,\phi) + \frac{1}{4\beta^2}}\right\} \operatorname{erfc}\left( \beta C(v_\theta, \phi) + \frac{1}{2\beta} \right)  \right\}
\end{aligned}
\end{equation}

where
\begin{displaymath}
 C(v_\theta, \phi) = v_\theta^2 + Z\phi - H_c = (H - H_c)|_{v_L = 0}
\end{displaymath}.

The density integral then reduces to

\begin{equation}
\begin{aligned}
n(A_\theta, \phi, r) = \frac{1}{2} e^{-\left(Z A_\theta r\right)^2} & \left\{
 e^{-Z\phi} \int_{-\infty}^{\infty} dv_\theta\, e^{-r^2\left(v_\theta^2 + 2Zv_\theta A_\theta\right)} e^{-v_\theta^2} \operatorname{erfc}\left(\beta \left\{v_\theta^2 + Z\phi - H_c \right\}  \right) \right. \\
&- \left. e^{\frac{1}{4\beta^2}} e^{-H_c} \int_{-\infty}^{\infty} dv_\theta\,
e^{-r^2 \left(v_\theta^2 + 2Zv_\theta A_\theta \right)} \operatorname{erfc}\left(\beta\left\{ v_\theta^2 + Z\phi - H_c \right\} + \frac{1}{2\beta}   \right)
\right\}
\end{aligned}
\end{equation}

We can approximate these integrals as follows:

\begin{equation}
\int_{-\infty}^{\infty} f(x) \operatorname{erfc}\left(g(x^2)\right) dx \approx 2 \int_{-v_-}^{v_-} f(x) dx + \int_{-v_+}^{-v_-} \left(1 - \frac{2g(x^2)}{\sqrt{\pi}}\right) f(x) dx +  \int_{v_-}^{v_+} \left(1 - \frac{2g(x^2)}{\sqrt{\pi}}\right) f(x) dx
\end{equation}

where
\begin{equation}
v_{\pm} = \Re\left({z_\pm}\right) \, \big| \, g(z_\pm^2) = \pm \frac{\sqrt{\pi}}{2}
\end{equation}

This yields the following expression for the density:

\begin{equation}
\begin{aligned}
n(A_\theta, \phi, r) =  e^{-\left(Z A_\theta r\right)^2} &  \left\{\Big. e^{-Z\phi}\left[ J(v_-) + I(-v_-) - I(-v_+) + I(v_+) - I(v_-) \right] \right. \\
&\left. -  e^{\frac{1}{4\beta^2}} e^{-H_c} \left[ J'(v'_-) + I'(-v'_-) - I'(-v'_+) + I'(v'_+) - I'(v'_-) \right]    \right\}
\end{aligned}
\end{equation}

where
\begin{equation}
\begin{aligned}
v_\pm = &\Re\left( \sqrt{ \pm \frac{\sqrt{\pi}}{2\beta} + H_c - Z\phi  }   \right) \\
v'_\pm = &\Re\left( \sqrt{ \pm \frac{\sqrt{\pi}}{2\beta} + H_c - Z\phi - \frac{1}{2\beta^2} }   \right) \\
\end{aligned}
\end{equation}

\begin{equation}
J(v) = \frac{\sqrt{\pi}}{\tilde{r}} e^{ \frac{(r^2 Z A_\theta)^2}{ \tilde{r}^2} }\left[
\operatorname{erf}\left( \frac{ \tilde{r}^2v + r^2 Z A_\theta }{\tilde{r}} \right)
- \operatorname{erf}\left( \frac{r^2 Z A_\theta - \tilde{r}^2v }{\tilde{r}} \right)
 \right]
\end{equation}

\begin{equation}
\begin{aligned}
I(v) = &\frac{\exp\left\{- \left( \tilde{r}^2 v^2 + 2r^2 Z A_\theta v \right)\right\} }{ 4\sqrt{\pi} \tilde{r}^5 }  \\
& \times \left\{\Bigg. \sqrt{\pi} \exp\left\{ \left(\tilde{r}^2 v + r^2 Z A_\theta\right)^2 / \tilde{r}^2 \right\} \operatorname{erf}\left(\frac{ \tilde{r}^2 v + r^2 Z A_\theta }{\tilde{r}}\right)\left[ 2\tilde{r}^2 \left( \sqrt{\pi} - 2 \beta \left(Z \phi - H_c \right) \right) - 2 \tilde{r}^2 \beta - \beta(2r^2ZA_\theta)^2  \right] \right. \\
& \left. \hspace{5mm} + 2 \tilde{r} \beta \left[ 2\tilde{r}^2 v + 2 r^2 Z A_\theta  \right]
 \Bigg. \right\}
\end{aligned}
\end{equation}

\begin{equation}
\begin{aligned}
I'(v') = &\frac{\exp\left\{- \left( {r}^2 v'^2 + 2r^2 Z A_\theta v' \right)\right\} }{ 4\sqrt{\pi} {r}^5 }  \\
& \times \left\{\Bigg. \sqrt{\pi} \exp\left\{ \left({r}^2 v' + r^2 Z A_\theta\right)^2 / {r}^2 \right\} \operatorname{erf}\left(\frac{ {r}^2 v' + r^2 Z A_\theta }{{r}}\right)\left[ 2{r}^2 \left( \sqrt{\pi} - 2 \beta \left(Z \phi - H_c \right) - \frac{1}{\beta} \right) - 2 {r}^2 \beta - \beta(2r^2ZA_\theta)^2  \right] \right. \\
& \left. \hspace{5mm} + 2 {r} \beta \left[ 2{r}^2 v' + 2 r^2 Z A_\theta  \right]
 \Bigg. \right\}
\end{aligned}
\end{equation}

\begin{equation}
J'(v') = \frac{\sqrt{\pi}}{r} e^{(r Z A_\theta)^2}\left[
\operatorname{erf}\left( rv' + rZA_\theta \right)
- \operatorname{erf}\left(r Z A_\theta - rv' \right)
 \right]
\end{equation}

and $\tilde{r} = \sqrt{r^2 + 1}$.
\end{widetext}

\end{document}